\documentclass[11pt]{article}
\usepackage{amsmath}
\usepackage{amsmath,amssymb}
\usepackage{geometry}
\usepackage{graphicx}
\usepackage{enumerate}
\usepackage{bm}
\usepackage{xcolor}
\usepackage{filecontents}

\usepackage{natbib}
\newtheorem{thm}{Theorem}
\newtheorem{prop}{Proposition}

\newcommand{\minn}[2]{\underset{#1}{\min}\left\{ #2 \right\}}

\newcommand{\argmax}[2]{\underset{#1}{\mathrm{argmax}}\left\{ #2 \right\}}
\def\minimize#1#2{  {\underset{#1}{\mathrm{minimize}}}\left\{#2\right\}}

\usepackage{mathtools}

\newcommand{\C}{\mathfrak{C}}

\newcommand{\R}{\mathbb{R}}

\newcommand{\cost}{\mathrm{Cost}}

\newcommand{\thj}{\hat\tau_{j}}
\newcommand{\thi}[1]{\hat\tau_{#1}}

\usepackage{xr}
\title{Testing for a Change in Mean After Changepoint Detection}

  \author{Sean Jewell \\
	Department of Statistics, University of Washington \\
	Paul Fearnhead \\
    Department of Mathematics and Statistics, Lancaster University \\
    Daniela Witten \\
    Departments of Statistics and Biostatistics, University of Washington
    }

\begin{document}
\maketitle

\begin{abstract}
While many methods are available to detect structural changes in a time series, few procedures are available to quantify the uncertainty of these estimates post-detection. In this work, we fill this gap by proposing a new framework to test the null hypothesis that there is no change in mean around an estimated changepoint. We further show that it is possible to efficiently carry out this framework in the case of changepoints estimated by binary segmentation and its variants, $\ell_{0}$ segmentation, or the fused lasso. Our setup allows us to condition on much less information than existing approaches, which yields higher powered tests. We apply our proposals in a simulation study and on  a dataset of chromosomal guanine-cytosine  content. These approaches are freely available in the \texttt{R} package \texttt{ChangepointInference}
at \texttt{https://jewellsean.github.io/changepoint-inference/}. \\

\end{abstract}

\section{Introduction}

Detecting structural changes in a time series is a fundamental problem in statistics, 
 with a variety of applications \citep{bai1998estimating, bai2003computation, muggeo2010efficient, schroder2013adaptive,futschik2014multiscale, xiao2019accurate,harchaoui2007catching, hotz2013idealizing}. A structural change refers to  the phenomenon that at a certain (unknown) timepoint $\tau$, the law of the data may change: that is, observations $y_{1}, \ldots, y_{T}$ are heterogeneous, in the sense that $y_{1}, \ldots, y_{\tau} \sim F$, whereas $y_{\tau+1}, \ldots, y_{T} \sim G$, for distribution functions $F \neq G$. In the presence of possible structural changes,  it is of interest not only to estimate the times at which these changes occur --- that is, the value of $\tau$ --- but also to conduct statistical inference on the estimated changepoints. 

In this paper, we consider the most common changepoint model,
\begin{align}
Y_{t} = \mu_{t} + \epsilon_{t}, \quad \epsilon_{t} \overset{\text{iid}}{\sim} \mathrm{N}(0, \sigma^{2}), \quad t=1,\ldots,T,
\label{eq:obs-model}
\end{align}
and assume that $\mu_1,\ldots,\mu_T$ is  piecewise constant, in the sense that  $\mu_{\tau_j+1}=\mu_{\tau_j + 2 } = \ldots = \mu_{\tau_{j+1}} \neq \mu_{\tau_{j+1}+1}$, for $j=0,\ldots,K-1$, and $\mu_{\tau_{K} + 1} = \mu_{\tau_{K} + 2} = \ldots = \mu_{\tau_{K+1}}$. 
Here $0 = \tau_{0} < \tau_{1} < \ldots < \tau_{K} < \tau_{K+1} = T$, and  $\tau_1,\ldots,\tau_K$ represent the true  changepoints.   
Changepoint detection refers to the task of estimating the changepoint locations ${\tau}_{1}, \ldots,{\tau}_{{K}}$, and possibly the number of changepoints $K$. A huge number of proposals for this task have been made in the literature; see \cite{truong2020selective} and \cite{fearnhead2020relating} for a comprehensive review. These proposals can be roughly divided into two classes.  One class  iteratively searches for one changepoint at a time \citep{vostrikova1981detection,olshen2004circular, fryzlewicz2014wild, badagian2015time, anastasiou2019detecting}; 
the canonical example of this approach is binary segmentation. Another class of proposals simultaneously estimates all changepoints by solving a single optimization problem \citep{auger1989algorithms, jackson2005algorithm, tibshirani2005sparsity, niu2012screening, killick2012optimal, haynes2017computationally, maidstone2017optimal, jewell2018exact, fearnhead2018detecting, hocking2018generalized, jewell2018fast}; examples include $\ell_0$ segmentation and the fused lasso. We review these approaches in Section~\ref{sec:background}. Although not a focus of our work,  changepoint estimation and inference have also been studied from a Bayesian perspective \citep{fearnhead2006exact,nam2012quantifying, ruanaidh2012numerical}. 

In the single changepoint setting, estimation and inference for the location of the changepoint  have been studied in the asymptotic \citep{hinkley1970inference, yao1987approximating, barry1987tests, bai1994least} and  non-asymptotic \citep{enikeeva2019high} settings. These approaches are typically extended to the multiple changepoint case by repeated application of tests for a single changepoint to sliding subsets of the data. 

In the multiple changepoint setting, the multiscale approach of \cite{frick2014multiscale} estimates the changepoints and provides confidence intervals for the changepoint locations and the unknown mean. However, this approach aims to control the probability of falsely detecting a change, and can lose power when there are many changes, particularly when they are hard to detect. Similarly, \cite{ma2016pairwise} produce asymptotically valid confidence intervals, but assume an asymptotic regime where all of the changepoints are detected with probability tending to one; this regime is unrealistic in many settings. 

To overcome these issues, \cite{li2016fdr} develop a multiscale procedure that controls the false discovery rate of detections. But their method uses a very weak definition of a ``true changepoint''. In extreme cases, this could include an estimated changepoint that is almost as far as $T/2$ observations from an actual changepoint. 

Non-parametric approaches to estimate multiple changepoints, such as moving-sum or scan statistics, have also been proposed \citep{bauer1980extension, huvskova1990asymptotics, chu1995mosum}.  \cite{eichinger2018mosum} recently showed consistency for the number and locations of changepoints, and established rates of convergence for moving-sum statistics. 

Despite the huge literature on estimation and inference in changepoint detection problems, there remains a gap between the procedures used by practitioners to estimate changepoints and the statistical tools to assess the uncertainty of these estimates:%. In particular, existing tools are of limited use due to the fact that: 

\begin{enumerate}
 \item Much of the theory for changepoint detection, especially in the multiple changepoint setting, focuses on specialized estimation  procedures that are designed to facilitate inference. Therefore, these results are not directly applicable to the procedures commonly used by data analysts to estimate changepoints  in practice. 
 \vspace{-2mm}
 \item Classical techniques to test for a single changepoint give (mostly) asymptotic results, which involve complicated limiting distributions that do not directly apply to the multiple changepoint setting.
 \vspace{-2mm}
 \item Earlier works (mostly) provide confidence statements for the location of the changepoint. However, downstream analyses often rely on the size of the shift in mean, and not its precise location. 
\end{enumerate}

To address these limitations, we consider testing the null hypothesis that there is no change in mean around an estimated changepoint. 
Our interest lies not in determining whether there is a change in mean at a precise location, but rather, whether there is a change in mean nearby. This is a challenging task, since  we must account for the  fact that the changepoint was estimated from the data --- and thus that the null hypothesis was chosen using the data --- when deriving the null distribution for a test statistic. A recent promising line of work was introduced by \cite{hyun2016exact} and \cite{hyun2018post}, who develop valid tests for a change in mean associated with changepoints  estimated with the generalized lasso or binary segmentation, respectively.
Their work leverages recent results for selective inference in the regression setting \citep{fithian2014optimal, fithian2015selective, tibshirani2016exact, lee2016exact,  tian2018selective}. In greater detail, they compute the probability of observing such a large change in mean associated with an estimated changepoint, conditional on the fact that the changepoint was estimated from the data, as well as some additional quantities required for computational tractability. However, the fact that they condition on much more information than is used to choose the null hypothesis that is tested leads to a substantial reduction in power, as pointed out by \cite{fithian2014optimal}, \cite{lee2016exact}, and \cite{liu2018more}. 

In this paper, we consider testing for a change in mean associated with an estimated changepoint, while conditioning on far less information than \cite{hyun2016exact} and \cite{hyun2018post}. In effect, we conduct local conditioning,  as opposed to the global conditioning needed in \cite{hyun2018post}. Moreover, we develop a test for a change in mean associated with changepoints  detected via $\ell_{0}$ segmentation, rather than only fused lasso and binary segmentation. Both of  these advances lead to more powerful procedures for testing for the presence of changepoints. We develop this framework in detail for the change-in-mean model, but the general ideas can be applied more widely.

The rest of this paper is organized as follows. In Section~\ref{sec:background}, we review the relevant literature on changepoint detection and inference. In Section~\ref{sec:selective-inference}, we introduce a framework for inference in changepoint detection problems, which allows us to test for a change in mean associated with a changepoint estimated on the same dataset. In Sections~\ref{sec:bs-characterization} and \ref{sec:l0-characterization}, we develop efficient algorithms that allow us to instantiate this framework in the special cases of  binary segmentation \citep{vostrikova1981detection} and its variants  \citep{olshen2004circular, fryzlewicz2014wild}, and $\ell_0$ segmentation \citep{killick2012optimal, maidstone2017optimal}; the case of the fused lasso \citep{tibshirani2016exact} is straightforward and addressed in the Supplementary Materials. Our framework is an improvement over the existing approaches for inference on the changepoints estimated using binary segmentation and its variants and the fused lasso; it is completely new in the case of $\ell_{0}$ segmentation. After a preprint of this work appeared \citep{jewell2019testing}, another research group developed a less efficient dynamic programming approach to assess the uncertainty in changepoints estimated from $\ell_{0}$ segmentation \citep{duy2020computing}. 
In Section~\ref{sec:experiments}, we present a comparison to some recent proposals from the 
literature  in a simulation study. In Section~\ref{sec:real-data-example}, we show that our procedure leads to additional discoveries versus existing methods on a dataset of chromosomal guanine-cytosine (G-C) content. Extensions are in Section~\ref{sec:discussion}, and some additional details are deferred to the Supplementary Materials. 

The \texttt{R} package \texttt{ChangepointInference}, along with code and data to reproduce all figures, can be found at  \texttt{https://jewellsean.github.io/changepoint-inference}. 

\section{Background}
\label{sec:background}
\subsection{Changepoint detection algorithms} 
\label{sec:background-detection}

\subsubsection{Binary segmentation and its variants} 
\label{sec:background-bs}
Binary segmentation \citep{vostrikova1981detection} and its variants  \citep{olshen2004circular, fryzlewicz2014wild} search for changepoints by solving a sequence of local optimization problems. For the change-in-mean problem, these use the cumulative sum (CUSUM) statistic
\begin{align}
g^{\top}_{(s, \tau, e)}y := \sqrt{\frac{1}{\frac{1}{|e - \tau|} + \frac{1}{|\tau+1-s|}}}(\bar{y}_{(\tau+1):e} - \bar{y}_{s:\tau}),
\label{eq:g-contrast}
\end{align}
defined through a contrast $g_{(s, \tau, e)}\in\R^{T}$, which summarizes the evidence for a change at $\tau$ in the data $y_{s:e}:= (y_{s}, \ldots, y_{e})$ by the difference in the empirical mean of the data before and after $\tau$ (normalized to have the same variance for all $\tau$). In \eqref{eq:g-contrast}, the notation $\bar{y}_{a:b}$ represents the sample mean of $(y_a,\ldots,y_b)$. 

In binary segmentation \citep{vostrikova1981detection}, the set of estimated changepoints is simply the set of local CUSUM maximizers: the first estimated changepoint maximizes the CUSUM statistic over all possible locations,
$\thi{1} = \argmax{\tau \in [1:(T-1)]}{|g^{\top}_{(1, \tau, T)}y|}.$
Subsequent changepoints are estimated at the location that maximizes the CUSUM statistic when we consider regions of the data between previously estimated changepoints. For example, the second estimated changepoint is
$\thi{2} = \argmax{\tau \in [1:(T-1)]\setminus \thi{1}}{|g^{\top}_{(1, \tau, \thi{1})}y| 1_{(1 \leq \tau < \thi{1})} + |g^{\top}_{(\thi{1}, \tau, T)}y| 1_{(\thi{1} < \tau < T)}}.$
 We continue in this manner until a stopping criterion is met. 
%Variants of this procedure have been proposed to improve performance \citep{olshen2004circular, fryzlewicz2014wild}.

\subsubsection{Simultaneous estimation of changepoints}
\label{sec:l0-opt-chgpt}

As an alternative to sequentially estimating changepoints, we can simultaneously estimate all changepoints by minimizing a penalized cost 
that trades off fit to the data against the number of changepoints \citep{killick2012optimal, maidstone2017optimal}, i.e.
\begin{align}
\minimize{\substack{0=\tau_{0} < \tau_{1} < \cdots < \tau_{K} < \tau_{K+1} = T, \\ u_{0}, u_1,\ldots, u_{K}, K}}{\frac12 \sum_{k = 0}^{K} \sum_{t=\tau_{k} + 1}^{\tau_{k + 1}} \left(y_{t} - u_{k}\right)^{2} + \lambda K}.
\label{eq:penalized-mini-l2}
\end{align}
This is equivalent to solving an $\ell_{0}$ penalized regression problem
\begin{align}
\minimize{\mu\in\R^{T}}{\frac12 \sum_{t = 1}^{T} (y_{t} - \mu_{t})^{2} + \lambda \sum_{t = 1}^{T-1} 1_{(\mu_{t} \neq \mu_{t+1})}},
\label{eq:l0-mean-rss}
\end{align}
in the sense that the vector $\hat\mu$ that solves \eqref{eq:l0-mean-rss} 
satisfies $\{t: \hat\mu_t \neq \hat\mu_{t+1} \} = \{ \hat\tau_1,\ldots,\hat\tau_{\hat{K}} \}$, 
%is piecewise constant with breakpoints at $\thi{1}, \ldots, \thi{\hat{K}}$, 
 where $\thi{1}, \ldots, \thi{\hat{K}}$ are the changepoints that solve \eqref{eq:penalized-mini-l2}. The tuning parameter $\lambda$  specifies the improvement in fit to the data needed to add an additional changepoint.

Replacing the $\ell_{0}$ penalty in \eqref{eq:l0-mean-rss} with an $\ell_{1}$ penalty leads to the well-studied trend filtering or fused lasso optimization problem \citep{rudin1992nonlinear, tibshirani2005sparsity},
\begin{align}
\minimize{\mu\in\R^{T}}{\frac12 \sum_{t = 1}^{T} (y_{t} - \mu_{t})^{2} + \lambda \sum_{t = 1}^{T-1} |\mu_{t} - \mu_{t+1}|}.
\label{eq:l1-mean-rss}
\end{align}

\subsection{Existing methods for inference on changepoints post-detection} 
\label{sec:background-inference}

Suppose that we estimate some changepoints $\thi{1}, \ldots, \thi{\hat{K}}$, and then wish to quantify the evidence for these estimated changepoints. 
We might naively apply a standard $z$-test for the difference in mean around each estimated changepoint. However,  this approach is problematic, because it entails using the same data for testing that was used to estimate the changepoints, and  thus to select the hypotheses to be tested. In particular, the  $z$-statistic is not normally distributed under the null hypothesis. In the linear regression setting, \cite{tibshirani2016exact} and \cite{lee2016exact} have shown  that it is possible to select and test hypotheses based on the same set of data, provided that we condition on the output of the hypothesis selection procedure. 

\cite{hyun2016exact} and \cite{hyun2018post} extend these ideas to the changepoint detection setting. 
 For each changepoint $\thj$ estimated using either binary segmentation, its variants, or the fused lasso, \cite{hyun2018post} propose to test 
whether there is a change in mean around $\thj$. They construct the test statistic $\hat{d}_{j} \nu_{j}^{\top}Y$, where $\hat{d}_{j}$ is the  sign of the estimated change in mean at $\thj$, and $\nu_{j}$ is a $T$-vector of contrasts, defined as 
\begin{equation}
\nu_{j,t} = 
\begin{cases}
0 & \text{ if } t \leq \hat\tau_{j-1} \text { or } t > \hat\tau_{j+1}, \\
\frac{1}{\hat\tau_{j } - \hat\tau_{j-1}} & \text { if } \hat\tau_{j-1} < t \leq \hat\tau_j, \\
- \frac{1}{\hat\tau_{j+1}-\hat\tau_j} & \text { if }   \hat\tau_{j} < t \leq \hat\tau_{j+1}.
\end{cases}
 \label{eq:nu}
\end{equation}
They consider the null hypothesis 
$
H_{0}:\hat{d}_{j} \nu_{j}^{\top}\mu = 0
$
versus the one-sided alternative $H_{1}: \hat{d}_{j} \nu_{j}^{\top}\mu > 0$. Since both $\hat{d}_j$ and $\nu_{j}$ are  functions of the estimated changepoints, it is clear that valid inference requires somehow conditioning on the estimation process, in the spirit of \cite{tibshirani2016exact} and \cite{lee2016exact}. Define $\mathcal{M}(y)$ to be the set of  changepoints estimated from the data $y$, i.e., $\mathcal{M}(y) = \{ \thi{1}, \ldots, \thi{\hat{K}}\}$. Then, it is tempting to define the $p$-value as
%\begin{equation} 
$ \text{Pr}_{H_0} \left(  \hat{d}_{j} \nu_{j}^{\top} Y  \geq  \hat{d}_{j} \nu_{j}^{\top} y   \mid \mathcal{M}(Y) = \mathcal{M}(y)  \right)$.
%\label{eq:p}
%\end{equation}
However, this $p$-value is not immediately amenable to the selective inference framework proposed by \cite{tibshirani2016exact} and  \cite{lee2016exact}, which requires that the conditioning set be polyhedral; i.e., the conditioning set can be written as $\{Y : \bm{A}Y \leq b\}$ for a matrix $\bm{A}$ and vector $b$.
  Thus, in the case of binary segmentation, \cite{hyun2018post} condition on three additional quantities: (i) the order in which the estimated changepoints enter the model, $\mathcal{O}(Y) = \mathcal{O}(y)$; (ii) the sign of the change in mean due to the estimated changepoints, $\Delta(Y) = \Delta(y)=(\hat{d}_1,\ldots,\hat{d}_{\hat{K}})$; (iii) $\Pi_{\nu_{j}}^{\perp} Y = \Pi_{\nu_{j}}^{\perp} y$, where $\Pi_{\nu_{j}}^{\perp} = I - \nu_{j} \nu_{j}^{\top} / ||\nu_{j}||_{2}^{2}$ is the orthogonal projection matrix onto the subspace that is orthogonal to $\nu_{j}$. Conditioning on (i) and (ii) ensures that the conditioning set is polyhedral, whereas conditioning on (iii) ensures that the test statistic is a pivot. This leads to the $p$-value
\begin{equation} \small
 \text{Pr}_{H_0} \left(  \hat{d}_{j} \nu_{j}^{\top} Y  \geq  \hat{d}_{j} \nu_{j}^{\top} y   \mid \mathcal{M}(Y) = \mathcal{M}(y), \mathcal{O}(Y) = \mathcal{O}(y), \Delta(Y) = \Delta(y), \Pi_{\nu_{j}}^{\perp} Y = \Pi_{\nu_{j}}^{\perp} y  \right),
\label{eq:p-bs-extra}
\end{equation}
which can be easily computed because the conditional distribution of $\hat{d}_{j}\nu_{j}^{\top}Y$ is a Gaussian truncated to an interval. 
 For slightly different conditioning sets, \cite{hyun2018post}  show similar results for variants of binary segmentation and for the fused lasso.

Importantly, \cite{hyun2018post} choose the conditioning set in \eqref{eq:p-bs-extra} for computational reasons:  there is no clear statistical motivation for conditioning on $\mathcal{O}(Y) = \mathcal{O}(y)$ and $\Delta(Y) = \Delta(y)$. Furthermore, it might be possible to account for the fact that changepoints are estimated from  the data without conditioning on the full set $\mathcal{M}(Y) = \mathcal{M}(y)$. In fact, \cite{fithian2014optimal} argue that when conducting selective inference, it is better to condition on less information, i.e. to condition on $Y$ being in a larger set of possible data, 
since conditioning on more information reduces the Fisher information that remains in the conditional distribution of the data.

For this reason, in the regression setting, some recent proposals seek to increase the size of the conditioning set. \cite{lee2016exact} propose to condition on just the selected model, rather than on the selected model and the corresponding coefficient signs, by considering all possible configurations of the signs of the estimated coefficients. Unfortunately, this comes at a significant computational cost. Continuing in this vein, \cite{liu2018more} partition the selected variables into high value and low value subsets, and then condition on the former and the variable of interest.

In this paper, we develop new insights that allow us to test the null hypothesis that there is no change in mean at an estimated changepoint,  without restriction to the polyhedral conditioning sets pursued by \cite{hyun2018post}. Because we do not need to use the full conditioning set in \eqref{eq:p-bs-extra}, we obtain  higher-powered tests. Additionally, since we avoid conditioning on $\Delta(Y) = \Delta(y)$, we can consider two-sided tests  of 
\begin{align}
H_{0}: \nu^{\top}\mu = 0 \text { versus } H_{1}:  \nu^{\top}\mu \neq 0,
\label{eq:hyp-test}
\end{align}
rather than the one-sided tests considered by \cite{hyun2018post}. In \eqref{eq:hyp-test}, and for the remainder of this paper, we suppress the $j$ subscript on $\nu_j$ for notational convenience. Thus, the vector $\nu$ should be interpreted as shorthand for $\nu_j$.

It is natural to ask whether we can avoid the complications of selective inference and use alternative approaches that control the false discovery rate \citep{benjamini1995controlling, benjamini2001control, barber2015controlling, candes2018panning}. However, these alternatives are not suitable for the changepoint setting in the following sense. Often we do not want to know if a true changepoint is \emph{exactly} at $\thj$, but rather whether there is a true changepoint \emph{near} $\thj$; that is, we are willing to accept small estimation errors in the location of a changepoint. With a suitable choice of $\nu$ in \eqref{eq:hyp-test}, we can test whether there is a change in mean near $\thj$, where \emph{near} can be defined appropriately for a given application. By contrast, while knockoffs \citep{barber2015controlling} or a related approach could likely be used to test for a change in mean at a precise location, in our experience such approaches tend to have almost no power to detect modest changes in the mean, due to the large uncertainty in the precise location of the change.

\subsection{Toy example illustrating the cost of conditioning}
\label{sec:toy-motivate-example}

\begin{figure}[h!]
\centering
\includegraphics[width = 0.8\textwidth]{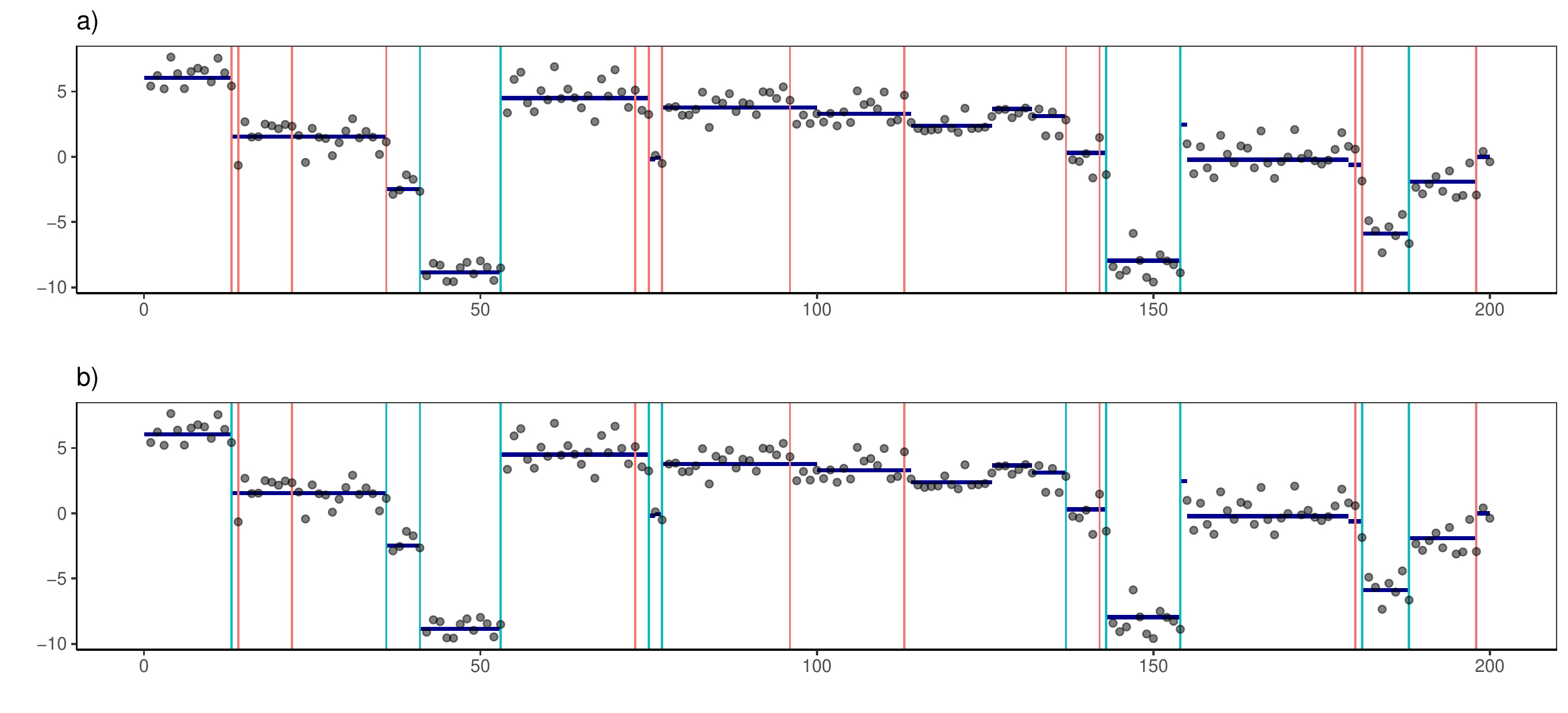}
\caption{The power of a test of \eqref{eq:hyp-test} critically depends on the size of the conditioning set. 
Observations (displayed in grey) were simulated from \eqref{eq:obs-model} with $\sigma = 1$ and $\mu_1,\ldots,\mu_T$ displayed in dark blue. Our proposed test of \eqref{eq:hyp-test} was conducted for each of the changepoints estimated via $19$-step binary segmentation.   Estimated changepoints for which the  $p$-value is less than $0.05$ are displayed in blue, and the remaining estimated changepoints are displayed in red. In panel (a), we conducted our proposed test by conditioning on $\mathcal{M}(Y) = \mathcal{M}(y), \mathcal{O}(Y) = \mathcal{O}(y), \Delta(Y) = \Delta(y),$ and $ \Pi_\nu^{\perp} Y = \Pi_\nu^{\perp} y$ (this is essentially the proposal of \cite{hyun2018post}). In panel (b), we conditioned on the much larger set  $\mathcal{M}(Y) = \mathcal{M}(y)$ and $ \Pi_\nu^{\perp} Y = \Pi_\nu^{\perp} y$.}
\label{fig:motivate}
\end{figure}

In this section, we demonstrate that the power of  a test of \eqref{eq:hyp-test} critically depends on the size of the conditioning set.  In Figure~\ref{fig:motivate}, we  consider two choices for the conditioning set.  In panel a), we condition on $\mathcal{M}(Y) = \mathcal{M}(y), \mathcal{O}(Y) = \mathcal{O}(y), \Delta(Y) = \Delta(y),$ and $ \Pi_\nu^{\perp} Y = \Pi_\nu^{\perp} y$: this is essentially the test proposed by \cite{hyun2018post}. In panel b) we condition on just $\mathcal{M}(Y) = \mathcal{M}(y)$ and $\Pi_\nu^{\perp} Y = \Pi_\nu^{\perp} y$. Observed data (grey points) are simulated according to  \eqref{eq:obs-model} with the true underlying mean displayed in blue. $19$-step binary segmentation is used to estimate changepoints, which are displayed as vertical lines, and are colored  based on whether the associated $p$-value is less than $0.05$ (blue) or greater than $0.05$ (red). In this example, \emph{conditioning on less information allows us to reject the null hypothesis when it is false more often (i.e., we obtain five additional true positives), without inflating the number of false positives}. 

With this toy example in mind, we turn to our proposal in the following section. It does not require polyhedral conditioning sets, and thus allows us to condition on much less information than previously possible.

\section{Two new tests with larger conditioning sets} 
\label{sec:selective-inference}

In this section, we consider testing a null hypothesis of the form \eqref{eq:hyp-test} using a much larger conditioning set than  used by \cite{hyun2018post}. Our approach is similar in spirit to the ``general recipe'' proposed in Section~6 of \cite{liu2018more}. We consider two possible forms of the contrast vector $\nu$ in Sections~\ref{sec:biggerconditioning} and \ref{sec:smallerconditioning}. 

\subsection{A test of no change in mean between neighboring changepoints} 

\label{sec:biggerconditioning}

We first consider testing the null hypothesis  \eqref{eq:hyp-test} for $\nu$ defined in \eqref{eq:nu}. 
In order to account for the fact that we estimated the changepoints, it is natural to condition 
  on all of the estimated changepoints, $\mathcal{M}(y) = \{\thi{1}, \ldots, \thi{\hat{K}}\}$. Thus, we define the $p$-value 
 \begin{equation}
 p \equiv \text{Pr}_{H_0} \left(  | \nu^{\top} Y | \geq | \nu^{\top} y |  \mid \mathcal{M}(Y) = \mathcal{M}(y)  , \Pi_\nu^{\perp} Y = \Pi_\nu^{\perp} y \right).
\label{eq:p2}
\end{equation}
As in \cite{hyun2018post}, we condition on $\Pi_\nu^{\perp} Y = \Pi_\nu^{\perp} y$ for technical reasons; see \cite{fithian2014optimal} for additional discussion. Roughly speaking, \eqref{eq:p2} asks: ``Out of all data sets yielding this particular set of changepoints, what is the probability, under the null that there is no true change in mean at this location, of observing such a large difference in mean between the segments on either side of $\hat\tau_j$?" % is as large as what is observed?'' 
Our next result reveals that computing \eqref{eq:p2} involves a univariate truncated normal distribution. Related results appear in \cite{tibshirani2016exact}, \cite{lee2016exact}, and \cite{liu2018more}. 
 \begin{thm}
\label{thm:pval}
 The $p$-value in \eqref{eq:p2} is equal to 
  \begin{equation}
 p = \text{Pr}\left(  | \phi | \geq | \nu^{\top} y | \mid \mathcal{M}(y'({\phi})) = \mathcal{M}(y)   \right),
 \label{eq:pfinal}
 \end{equation}
where  $\phi \sim N(0, \| \nu \|^2 \sigma^2)$ and where
\begin{align}
y'(\phi)= y - \frac{\nu \nu^{\top}y}{||\nu||_{2}^{2}}  + \frac{\nu \phi}{||\nu||_{2}^{2}}.
\label{eq:yphi}
\end{align}
 \end{thm}
In light of Theorem~\ref{thm:pval}, to evaluate \eqref{eq:p2} we must simply characterize the set 
\begin{align}
\mathcal{S} = \{\phi: \mathcal{M}(y'(\phi)) = \mathcal{M}(y) \};
\label{eq:set-s}
\end{align}
as we will see in Section~\ref{sec:intuition}, this is the set of perturbations of $y$ that result in no change to the estimated changepoints. 
 In Sections~\ref{sec:bs-characterization} and \ref{sec:l0-characterization}, we do exactly this in the case of binary and $\ell_0$ segmentation, respectively. We discuss the fused lasso in Section~\ref{sec:fused-lasso-characterization} of the Supplementary Materials.

\subsection{A test of no change in mean within a fixed window size}
\label{sec:smallerconditioning}

We now consider testing the null hypothesis \eqref{eq:hyp-test} with $\nu$ given by 
\begin{equation}
\nu_t = 
\begin{cases}
0 & \text{ if } t \leq \thj - h \text { or } t > \thj + h, \\
\frac{1}{h} & \text { if } \thj - h < t \leq \thj, \\
- \frac{1}{h} & \text { if }   \thj < t \leq \thj + h.
\end{cases}                                            
 \label{eq:nu-window}
\end{equation}
Thus, we are testing whether the mean in a window to the left of the $j$th estimated changepoint equals the mean in a window  to the right of the $j$th estimated changepoint, for a fixed window size $h>0$.    
When considering this null hypothesis, it makes sense to condition only on the $j$th estimated changepoint, 
 leading to a $p$-value defined as
 \begin{equation}
 p \equiv \text{Pr}_{H_0} \left(  | \nu^{\top} Y | \geq | \nu^{\top} y |  \mid \thj \in \mathcal{M}(Y)  , \Pi_\nu^{\perp} Y = \Pi_\nu^{\perp} y \right),
\label{eq:p2-thj}
\end{equation}
where once again, we condition on $ \Pi_\nu^{\perp} Y = \Pi_\nu^{\perp} y $ for technical reasons. 
Roughly speaking,  \eqref{eq:p2-thj} asks: ``Out of all  data sets yielding a changepoint at $\thj$, what is the probability, under the null that there is no true change in mean at this location, of observing such a large difference in mean between the windows of size $h$ on either side of $\thj$?''

The $p$-values in \eqref{eq:p2-thj} and \eqref{eq:p2} are calculated for slightly different null hypotheses: the null for \eqref{eq:p2-thj} is that there is no changepoint within a distance $h$ of the estimated changepoint, $\thj$. By contrast, \eqref{eq:p2} tests for no change in mean between the estimated changepoints immediately before and after $\thj$. Furthermore, \eqref{eq:p2-thj}  conditions on less information. We believe that in many applications, the null hypothesis assumed by \eqref{eq:p2-thj}  is more natural and informative, since it allows a practitioner to specify how accurately they want to detect changepoint locations, and it avoids rejecting the null due to changes that are arbitrarily far away from $\thj$. Moreover, the ability to condition on less information intuitively should lead to higher power. If required, the ideas used to calculate \eqref{eq:p2-thj}  could also be applied to test for the null hypothesis assumed by \eqref{eq:p2}, while conditioning on less information. We further investigate these issues in Sections~\ref{sec:experiments} and \ref{sec:smaller-conditioning-p2}.

Theorem~\ref{thm:pval} can be extended to show that \eqref{eq:p2-thj} is equal to
 \begin{align}
 p &=  \text{Pr}\left(  | \phi | \geq | \nu^{\top} y | \mid \thj \in \mathcal{M}(y'({\phi}))    \right),
 \label{eq:pfinal-thj}
 \end{align}
 where $\phi  \sim N(0, \| \nu \|^2 \sigma^2)$, and where $y'(\phi)$ was defined in \eqref{eq:yphi}. Thus, computing the $p$-value requires 
 characterizing the set 
\begin{align}
\mathcal{S} = \{\phi: \thj \in \mathcal{M}(y'(\phi)) \};
\label{eq:set-s-thj}
\end{align}
this is the set of perturbations of $y$ that result in estimating a changepoint at $\thj$.

We show in Sections~\ref{sec:bs-characterization} and \ref{sec:l0-characterization} that $\mathcal{S}$ can be efficiently characterized for binary and $\ell_0$ segmentation. We discuss the fused lasso in Section~\ref{sec:fused-lasso-characterization} of the Supplementary Materials.

\subsection{Intuition for $y'(\phi)$ and $\mathcal{S}$}
\label{sec:intuition}
To gain intuition for $y'(\phi)$ in \eqref{eq:yphi}, 
we consider  $\nu$ defined in \eqref{eq:nu} (similar results apply for $\nu$ defined in \eqref{eq:nu-window}). We see that  
\begin{equation}
 y'_t(\phi) \equiv 
\begin{cases}
y_t & \text{ if } t \leq \hat\tau_{j-1} \text { or } t > \hat\tau_{j+1},\\
y_t + \frac{\phi - \nu^{\top} y}{ 1  + \frac{\hat\tau_{j } - \hat\tau_{j-1} }{\hat\tau_{j+1}-\hat\tau_j } }  & \text { if } \hat\tau_{j-1} < t \leq \hat\tau_j, \\
y_t -  \frac{\phi - \nu^{\top} y }{ 1  + \frac{\hat\tau_{j+1 } - \hat\tau_{j} }{\hat\tau_{j}-\hat\tau_{j-1} } }   & \text{ if }  \hat\tau_{j} < t \leq \hat\tau_{j+1}.
\end{cases} \label{eq:a}
\end{equation}
Thus,  $y'_{t}(\phi)$ is equal to $y_{t}$ for $t\leq \thi{j-1}$ or $t > \thi{j+1}$, and otherwise equals the observed data perturbed by a function of $\phi$ around $\thj$. 
 In other words, we can view $y'(\phi)$ as a perturbation of the observed data $y$ by a quantity proportional to $\phi - \nu^{\top} y$, within some window of $\thj$. Furthermore,  $\mathcal{S} = \{\phi: \mathcal{M}(y'(\phi)) = \mathcal{M}(y) \} $ is the set of such perturbations that do not affect the set of estimated changepoints. 

\begin{figure}[t!]
\centering
\includegraphics[width = 0.85\textwidth]{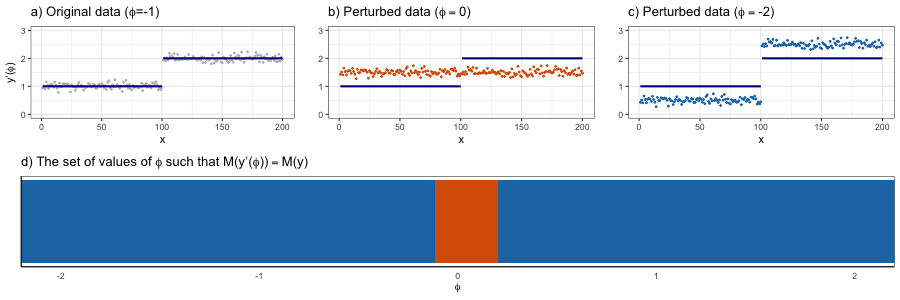}
\caption{ a) A simulated dataset with $\phi = \nu^{\top}y = -1$ is displayed in grey, and the true underlying mean is shown in blue. b)
The perturbed dataset $y'(\phi)$ is shown, with $\phi=\nu^{\top}y=0$. The perturbed dataset  does not have a change in mean at the $100$th timepoint, and so $1$-step binary segmentation does not detect a changepoint at that position. c) The perturbed dataset $y'(\phi)$ is shown, with $\phi=\nu^{\top}y=-2$. There is now a very pronounced change in mean at the $100$th timepoint, and so $1$-step binary segmentation does detect a changepoint at that position. 
d) Values of $\phi$ for which $\mathcal{M}(y'(\phi)) = \mathcal{M}(y)$  are shown in blue, and those for which $\mathcal{M}(y'(\phi)) \neq \mathcal{M}(y)$  are shown in red, for $\mathcal{M}$ given by 1-step binary segmentation.}
\label{fig:perturbed-data}
\end{figure}

Figure~\ref{fig:perturbed-data} illustrates the intuition behind $y'(\phi)$ in a simulated example with a change in mean at the $100$th position, and where $\phi=\nu^{\top}y = -1$. In panel a), the observed data are displayed. Here, $1$-step binary segmentation estimates $\thi{1} = 100$. In panel b), the observed data are perturbed using $\phi=0$ so that $1$-step binary segmentation no longer estimates a changepoint at the $100$th position. Conversely, in panel c), the data are perturbed using $\phi=-2$ to exaggerate the change at timepoint 100; $1$-step binary segmentation again estimates a changepoint at the $100$th position. Hence, for 1-step binary segmentation, $-1$ and $-2$ are in $\mathcal{S} = \{\phi: \mathcal{M}(y'(\phi)) = \mathcal{M}(y) \} $, but $0$ is not.  The procedure from Section~\ref{sec:bs-characterization} for efficiently characterizing $\mathcal{S}$ gives $\mathcal{S} = \{\phi: \mathcal{M}(y'(\phi)) = \mathcal{M}(y) \} = (-\infty, -0.2] \cup [0.2, \infty)$; see panel d) of Figure~\ref{fig:perturbed-data}.

%In  Sections~\ref{sec:bs-characterization} and \ref{sec:l0-characterization}, and in Section~\ref{sec:fused-lasso-characterization} of the Supplementary Materials, we develop procedures to characterize $\mathcal{S}$ in the cases of binary segmentation and its variants, $\ell_0$ segmentation, and the fused lasso, respectively. 

\section{Efficient characterization of \eqref{eq:set-s} and \eqref{eq:set-s-thj} for binary segmentation and its variants}
\label{sec:bs-characterization}

We now turn our attention to computing the set \eqref{eq:set-s} for $k$-step binary segmentation; \eqref{eq:set-s-thj} is detailed in Section~\ref{sec:basic-set-s-character-bs-2} of the Supplementary Materials.
%Extensions to variants of binary segmentation proposed in \cite{olshen2004circular} and \cite{fryzlewicz2014wild} are briefly discussed in Section~\ref{sec:extensions}.  
We begin by paraphrasing Proposition~\ref{prop:hyun} of \cite{hyun2018post}. 

\begin{prop}[Proposition~1 of Hyun et al., (2021)]
\label{prop:hyun}
 The set of $y$ for which $k$-step binary segmentation yields a given set of estimated changepoints, orders, and signs is polyhedral, and takes the form $\{y : \bm{\Gamma} y \leq 0 \}$ for a $k(2T - k - 3) \times T$ matrix $\bm{\Gamma}$, which is a function of the estimated changepoints, orders, and signs. 
\end{prop}
Recall from Section~\ref{sec:background-inference} that $\mathcal{M}(y)$, $\mathcal{O}(y)$, and $\Delta(y)$ are the locations, orders, and signs of the changepoints estimated from   $k$-step binary segmentation applied to  data $y$. 
%We now present a new result. 

\begin{prop}
\label{prop:s-decomposition}
The set $\{\phi : \mathcal{M}(y'(\phi)) = m, \mathcal{O}(y'(\phi)) = o, \Delta(y'(\phi)) = d \}$ is an interval. Furthermore,
the set $\mathcal{S}$  in \eqref{eq:set-s} is the union of such intervals,
\begin{align}
\mathcal{S} = \{\phi : \mathcal{M}(y'(\phi)) = \mathcal{M}(y) \} &= \bigcup_{i \in \mathcal{J}} [a_{i}, a_{i + 1}],
\label{eq:s-decomposition}
\end{align}
for an index set $\mathcal{J}$. 
%Suppose that a given value of $\phi$ results in estimated changepoints $\mathcal{M}(y'(\phi)) = \mathcal{M}(y)$; then Define the set
%$\mathcal{I}$ to be the order
%
 Let $\mathcal{I}$ denote  the set of  orders and signs of the changepoints that can be obtained via a perturbation of $y$ that yields changepoints $\mathcal{M}(y)$: that is, 
%where $\mathcal{J}$ is an index set, and $|\mathcal{J}|$ is the number of elements in the set
\begin{align}
\label{eq:num-elements-i}
\mathcal{I} := \left\{ (o, d) : \exists \phi \in \R \text{ such that }  \mathcal{M}(y'(\phi)) =  \mathcal{M}(y), \mathcal{O}(y'(\phi))=o, \Delta(y'(\phi))=d  \right\}.
\end{align}
 Then, $|\mathcal{J}|=|\mathcal{I}|$, i.e., the two sets have the same cardinality.
\end{prop}
Importantly, $|\mathcal{J}| = | \mathcal{I} | \ll 2^{k}k!$, which is the total number of possible orders and signs for the $k$ changepoints. 
To simplify notation in \eqref{eq:s-decomposition},  we have used the convention that if $a_i=-\infty$ then $[a_i,a_{i+1}]$ should be interpreted as $(a_i,a_{i+1}]$, and similarly if $a_{i+1}=\infty$ then $[a_i, a_{i+1}]$ should be interpreted as 
$[a_i, a_{i+1})$.

Proposition~\ref{prop:efficient-intervals} guarantees that Proposition~\ref{prop:s-decomposition} is of practical use.

\begin{prop}
\label{prop:efficient-intervals}
$\bigcup_{i\in\mathcal{J}} [a_{i}, a_{i+1}]$ defined in \eqref{eq:s-decomposition} can be efficiently computed. 
\end{prop}
Proposition~\ref{prop:efficient-intervals} follows from a simple argument. We first run $k$-step binary segmentation on the data $y$ to obtain estimated changepoints $\mathcal{M}(y)$, orders $\mathcal{O}(y)$, and signs $\Delta(y)$. We then apply the first statement in Proposition~\ref{prop:s-decomposition}  to obtain the interval $[a_0, a_1] = \{\phi : \mathcal{M}(y'(\phi)) = \mathcal{M}(y), \mathcal{O}(y'(\phi)) = \mathcal{O}(y), \Delta(y'(\phi)) = \Delta(y) \}$. By construction, $[a_0,a_1] \subset \mathcal{S}$.  The set $\mathcal{J}$ indexes the intervals comprising the set $\mathcal{S}$; therefore, we set $\mathcal{J} = \{0\}$. 

Next, for some small $\eta>0$, we apply the first statement in Proposition~\ref{prop:s-decomposition} with $m = \mathcal{M}(y'(a_1+\eta))$, $o = \mathcal{O}(y'(a_1+\eta))$, and $d = \Delta(y'(a_1+\eta))$ to obtain the interval $[a_1, a_2] = \{\phi : \mathcal{M}(y'(\phi)) = m, \mathcal{O}(y'(\phi)) = o, \Delta(y'(\phi)) = d \}$. 
(If the left endpoint of this interval does not equal $a_1$, then we must use a smaller value of $\eta$.) We then check whether $\mathcal{M}(y'(a_1+\eta))=\mathcal{M}(y)$. If so, then $[a_1, a_2] \subset \mathcal{S}$ and we set $\mathcal{J}$ equal to $\mathcal{J} \cup \{1\}$; if not, then $\mathcal{J}$ remains unchanged. Next, we apply the first statement of Proposition~\ref{prop:s-decomposition} with $m = \mathcal{M}(y'(a_2+\eta))$, $o = \mathcal{O}(y'(a_2+\eta))$, and $d = \Delta(y'(a_2+\eta))$ to obtain the interval $[a_2,a_3]$.
 We then determine whether $[a_2,a_3] \subset \mathcal{S}$; if so, then we set $\mathcal{J}$ equal to $\mathcal{J} \cup \{2\}$, and if not, then $\mathcal{J}$ remains unchanged. We continue in this way until we reach an interval containing $\infty$. We then repeat this process in the other direction, applying the first statement of Proposition~\ref{prop:s-decomposition} with $m = \mathcal{M}(y'(a_0-\eta))$, $o = \mathcal{O}(y'(a_0-\eta))$, and $d = \Delta(y'(a_0-\eta))$,  determining whether the resulting interval $[a_{-1}, a_{0}]$ belongs to $\mathcal{S}$, and updating $\mathcal{J}$ accordingly. We continue until  we arrive at an interval containing $-\infty$. 

Proposition~\ref{prop:early-stopping} shows that this procedure can be stopped early in order to  obtain conservative $p$-values, while substantially reducing computational costs.

\begin{prop}
\label{prop:early-stopping}
Let $\tilde{\mathcal{S}}$ be defined as the set 
\begin{align*}
\tilde{\mathcal{S}} = (-\infty, a_{-r}] \cup \left( \bigcup_{i \in \mathcal{J} \cap \{-r, \ldots, r'\}} [a_{i}, a_{i+1}] \right) \cup [a_{r' + 1}, \infty),
\end{align*}
for some $r$ and $r'$ such that $a_{-r} \leq -|\nu^{\top}y|$ and $a_{r'+1} \geq |\nu^{\top}y|$. Then the $p$-value obtained by conditioning on $\{ \phi \in \tilde{\mathcal{S}}\}$ exceeds the $p$-value obtained by conditioning on $\{ \phi \in \mathcal{S}\}$:
\begin{align*}
\text{Pr}(|\phi| \geq |\nu^{\top}y | \mid \phi \in \tilde{\mathcal{S}}) &\geq \text{Pr}(|\phi| \geq |\nu^{\top}y | \mid \phi \in \mathcal{S}).
\end{align*}
\end{prop}

Section~\ref{sec:efficient-bs-supp} of the Supplementary Materials contains proofs of Propositions~\ref{prop:s-decomposition} and \ref{prop:early-stopping}. In that section, we also show that Propositions~\ref{prop:s-decomposition} and \ref{prop:efficient-intervals} can be easily modified to characterize \eqref{eq:set-s-thj}. Section~\ref{sec:fused-lasso-characterization} of the Supplementary Materials contains a straightforward modification of this procedure to characterize \eqref{eq:set-s} and \eqref{eq:set-s-thj} in the case of the fused lasso.

It turns out that all of the ideas developed in this section for binary segmentation can be directly applied to the circular binary segmentation proposal of \cite{olshen2004circular} and the wild binary segmentation 
proposal of \cite{fryzlewicz2014wild}. In particular, it is shown in the Supplementary Materials of \cite{hyun2018post} that a result almost identical to Proposition~\ref{prop:hyun} holds for these two variants of binary segmentation, for a different matrix $\bf \Gamma$. This means that Propositions~\ref{prop:s-decomposition}--\ref{prop:early-stopping} follow directly.

We have assumed that $k$, the number of steps of binary segmentation, is pre-specified.  \cite{hyun2018post} showed that a stopping rule based on the Bayesian information criterion yields a  polyhedral conditioning set. Hence, we could  extend the ideas in this section to select $k$ adaptively. However, as shown by \cite{hyun2018post}, this approach requires conditioning on  additional information, and thereby results in a loss of power. 

\section{Efficient characterization of \eqref{eq:set-s} and \eqref{eq:set-s-thj} for $\ell_0$ segmentation} 
\label{sec:l0-characterization}

In this section, we develop an efficient algorithm to analytically characterize $\mathcal{S}$ in \eqref{eq:set-s} for the $\ell_0$ segmentation problem \eqref{eq:l0-mean-rss} with a fixed value of $\lambda$; Section~\ref{sec:set-s-thj-l0} of the Supplementary Materials considers $\mathcal{S}$ in  \eqref{eq:set-s-thj}. 
Recall that in the context of $\mathcal{S}$ in \eqref{eq:set-s}, $y'(\phi)$ is defined in \eqref{eq:yphi} and $\nu$ is defined in \eqref{eq:nu}.

%In particular, we wish to determine all values of $\phi$ such that $\phi \in \mathcal{S}$, without checking each value of $\phi$ individually. 
 Roughly speaking, we show that it is possible to write \eqref{eq:set-s} in terms of the cost to segment the perturbed data $y'(\phi)$. To compute the necessary cost functions, we derive recursions  similar to those in  \cite{rigaill2015pruned} and \cite{maidstone2017optimal}. However, these recursions involve functions of two variables, rather than one. Consequently, fundamentally different techniques are required for efficient computation. % to avoid a computational cost that increases exponentially.% in $h$.

\subsection{Recharacterizing $\mathcal{S}$ in \eqref{eq:set-s} in terms of $C(\phi)$ and $C'(\phi)$}

Let $\hat{K}$ denote the number of estimated changepoints resulting from $\ell_0$ segmentation \eqref{eq:l0-mean-rss} on the  data $y$ with fixed tuning parameter value $\lambda$, and let $\hat\tau_1 <\ldots < \hat\tau_{\hat K}$ denote the positions of those estimated changepoints; for notational convenience, let $\thi{0} \equiv 0$ and $\thi{\hat K + 1} \equiv T$. % Recall the definition of $y'(\phi)$ in \eqref{eq:yphi} and the definition of $\nu$ in \eqref{eq:nu}. 
For a given value of $\phi$, $\mathcal{M}(y'(\phi)) = \mathcal{M}(y) $ if and only if the cost of $\ell_0$ segmentation of the data $y'(\phi)$ with the changepoints restricted to occur at  $\hat\tau_1,\ldots,\hat\tau_{\hat K}$, 
\begin{equation}
C(\phi) = \min_{u_{0}, u_1,\ldots, u_{\hat{K}}} \left\{ \frac12 \sum_{k = 0}^{\hat{K}}  \sum_{t = \thi{k} + 1}^{\thi{k}} (y_{t}'(\phi) - u_{k})^{2}  + \lambda \hat K \right\},
\label{eq:Cphi}
\end{equation}
 is no greater than the cost of $\ell_0$ segmentation of $y'(\phi)$,
 \begin{equation}
 C'(\phi) = \min_{\substack{0=\tau_{0} < \tau_{1} < \cdots < \tau_{K} < \tau_{K+1} = T, \\ u_{0}, u_1,\ldots, u_{K}, K}} \left\{ \frac12 \sum_{k = 0}^{K}  \sum_{t = \tau_{k} + 1}^{\tau_{k + 1}} (y_{t}'(\phi) - u_{k})^{2}  + \lambda K \right\}.
 \label{eq:Cprimephi}
 \end{equation}  
In other words, 
\begin{equation}
\mathcal{S}=\{\phi: C(\phi) \leq C'(\phi)\}.
\label{eq:set-S2}
\end{equation} 
The following result follows from the fact that \eqref{eq:yphi} and \eqref{eq:nu}  imply that for all $j=0,\ldots,\hat{K}$, there exists a constant $c_j$ such that $y_t'(\phi) = y_t + c_j$ for all $t=\thi{j}+1,\ldots,\thi{j+1}$. %, for some constant $c_j$. 
% $y_{\thi{j}+1}'(\phi)= y_{\thi{j}+2}'(\phi)=\ldots=y_{\thi{j+1}}'(\phi)$ for all $j=0,\ldots,\hat{K}$.
 \begin{prop}
 $C(\phi)$ is a constant function of $\phi$. That is, $C(\phi)=C(\phi')$ for all $\phi$ and $\phi'$. 
  \label{prop:c-phi-constant-cost}
 \end{prop}

% Therefore, by inspection of \eqref{eq:Cphi}, $C(\phi)$ does not depend on the value of $\phi$.  

 Proposition~\ref{prop:c-phi-constant-cost} implies that $C(\phi)$ is easy to calculate: we just compute it for a single value of $\phi$, e.g. $\phi=\nu^T y$. 
  Hence, to characterize $\mathcal{S}$ using \eqref{eq:set-S2}, it remains to calculate $C'(\phi)$, i.e., to perform $\ell_0$ segmentation on $y'(\phi)$. In the interest of computational tractability, we need a single procedure that works for all values of $\phi$ simultaneously, rather than (for instance) having to repeat the procedure for values of $\phi$ on a fine grid. 

Let $\cost(y'_{1:\thj}(\phi); u)$ be the cost of segmenting $y'_{1:\thj}(\phi)$ with $\mu_{\thj} = u$. Then  $C'(\phi)$ can be decomposed into the cost of segmenting the data $y'(\phi)$ with a changepoint at $\thj$, 
 \begin{align}
C_{\thj}'(\phi) = \minn{u}{\cost(y'_{1:\thj}(\phi); u)} + \minn{u'}{\cost(y'_{T:(\thj+1)}(\phi); u')} + \lambda,
\label{eq:c-phi-thj}
\end{align}
and the cost of segmenting the data $y'(\phi)$ without a changepoint at $\thj$, 
 \begin{align}
C_{\neg\thj}'(\phi) = \minn{u}{\cost(y'_{1:\thj}(\phi); u) + \cost(y'_{T:(\thj+1)}(\phi); u)}.
\label{eq:c-phi-neg-thj}
\end{align}
 Combining \eqref{eq:c-phi-thj} and \eqref{eq:c-phi-neg-thj}, we have
\begin{align}
C'(\phi) = \minn{}{C_{\thj}'(\phi), C_{\neg\thj}'(\phi)}.
\label{eq:cprimephi}
\end{align}
Next, we will show that it is possible to analytically calculate $\cost(y_{1:\thj}'({\phi}); u)$ as a function of the perturbation, $\phi$, and the mean at the $\thj$th timepoint, $u$. A similar approach can be used to compute $\cost(y_{T:(\thj + 1)}'({\phi}); u)$.

\subsection{Analytic computation of $\cost(y_{1:\thj}'({\phi}); u)$}
\label{sec:analytic-computation-of-cost}

 We first note that $\cost(y_{1:s}; u)$, the cost of segmenting $y_{1:s}$ with $\mu_{s} = u$, can be efficiently computed \citep{rigaill2015pruned, maidstone2017optimal}. The cost at the first timepoint is simply $\cost(y_{1}; u) = \frac12(y_{1} - u)^{2}$. For any $s > 1$ and for all $u$, 
\begin{align}
\cost(y_{1:s}; u) = \min\left\{ \cost(y_{1:(s-1)};u), \minn{u'}{\cost(y_{1:(s-1)};u')} + \lambda	\right\} + \frac12(y_{s} - u)^{2}.
\label{eq:functional-updates}
\end{align}
For each $u$, this recursion encapsulates two possibilities: (i) there is no changepoint at  the $(s-1)$st timepoint, and the optimal cost is equal to the previous cost plus the cost of a new data point, $\cost(y_{1:(s-1)};u) + \frac12(y_{s} - u)^{2}$; (ii) there is a changepoint at the $(s-1)$st timepoint, and the optimal cost is equal to the optimal cost of segmenting up to $s-1$ plus the penalty for adding a changepoint at $s-1$ plus the cost of a new data point,  $\minn{u'}{\cost(y_{1:(s-1)};u')} + \lambda	+ \frac12(y_{s} - u)^{2}$. The resulting cost functions $\cost(y_{1}; u), \ldots, \cost(y_{1:T}; u)$ can be used to determine the exact solution to \eqref{eq:l0-mean-rss}. 

At first blush, the recursion in \eqref{eq:functional-updates} appears to be intractable due to the fact that, naively, $\cost(y_{1:s}; u)$ needs to updated for each value of $u\in\R$. However,  \cite{rigaill2015pruned} and \cite{maidstone2017optimal} show that these updates can be performed by efficiently manipulating piecewise quadratic functions of $u$, without needing to explicitly consider individual values of $u$, using a procedure that they call \textit{functional pruning}.  

It turns out that many of the computations made in the recursion \eqref{eq:functional-updates} can be reused in the calculation of $\cost(y_{1:\thj}'({\phi}); u)$. In particular, we note that from \eqref{eq:yphi} and \eqref{eq:nu}, $y'_{s}(\phi)=y_{s}$ for all $s \notin \{ \thi{j-1} +1, \ldots, \thi{j+1}\}$, and therefore, $\cost(y_{1:\thi{j-1}}'({\phi}); u) = \cost(y_{1:\thi{j-1}}; u)$. As a result, we only require a new algorithm to efficiently compute $\cost(y_{1:(\thi{j-1} + 1)}'({\phi}); u), \ldots,\cost(y_{1:\thj}'({\phi}); u)$. We now show that 
%will see that these are piecewise quadratic functions of two variables; therefore, developing functional pruning recursions similar to the one-dimensional recursions of \eqref{eq:functional-updates} is fundamentally more difficult. Nonetheless, in Theorem~\ref{thm:2d-pruning} we show that 
for $s= \thi{j-1}+1, \ldots, \thj$, $\cost(y_{1:s}'({\phi}); u)$  is the pointwise minimum over a set $\C_{s}$ of piecewise quadratic functions of $u$ and $\phi$ that can be efficiently computed. 

\begin{thm}
For $\thi{j-1} < s \leq \thj$, 
\begin{equation}
\cost(y_{1:s}'({\phi}); u) = \min_{f \in \C_{s}} f(u, \phi), \label{eq:thm2}
\end{equation}
where $\left\{ f(u, \phi) \right \}_{f\in \C_{s}}$ is a collection of $s - \thi{j-1} + 1$ piecewise quadratic functions of $u$ and $\phi$ constructed recursively from $\thi{j-1} + 1$ to $s$, and where $\C_{\thi{j-1}} = \{ \cost(y_{1:\thi{j-1}}; u) \}$. Furthermore, the set $\C_{\thj}$ can be computed in $\mathcal{O}((\thj - \thi{j-1})^{2})$ operations.
 \label{thm:2d-pruning}
\end{thm}
Section~\ref{sec:proof-thm}  of the Supplementary Materials contains a proof of Theorem~\ref{thm:2d-pruning}.      % and timing results, respectively. 

\subsection{Computing $C'(\phi)$ based on $\cost(y_{1:\thj}'({\phi}); u)$ and $\cost(y_{T:(\thj + 1)}'({\phi}); u)$}
\label{sec:compute-cprimephi}

Recall from \eqref{eq:cprimephi} that $C'(\phi)$ is the minimum of $C_{\thj}'(\phi)$ and $C_{\neg\thj}'(\phi)$, in \eqref{eq:c-phi-thj} and \eqref{eq:c-phi-neg-thj}, respectively. 
We now show how to compute $C_{\thj}'(\phi)$. 

We apply Theorem~\ref{thm:2d-pruning} to build the set $\C_{\thj}$, and recall from \eqref{eq:thm2} that $\cost(y_{1:\thj}'({\phi}); u) = \min_{f \in \C_{\thj}} f(u, \phi)$. Additionally, we define $\tilde{\C}_{\thi{j+1} + 1} = \{ \cost(y_{T:(\thi{j+1} + 1)}; u) \}$, and build $\tilde{\C}_{\thi{j+1}}, \ldots, \tilde{\C}_{\thj + 1}$ 
such that $\cost(y_{T:(\thj +1)}'(\phi); u) = \min_{f \in \tilde{\C}_{\thj + 1}} f(u, \phi)$,
 using a modification of Theorem~\ref{thm:2d-pruning} that accounts for the reversal of the timepoints.  
%By \eqref{eq:thm2}, $\cost(y_{1:\thj}'({\phi}); u) = \min_{f \in \C_{\thj}} f(u, \phi)$ and $\cost(y_{T:(\thj+1)}'({\phi}); u) = \min_{f \in \tilde{\C}_{\thj + 1}} f(u, \phi)$. 
Plugging into \eqref{eq:c-phi-thj}, 
\begin{align}
C_{\thj}'(\phi) 
 &= \minn{u}{\minn{f \in \C_{\thj}}{ f(u, \phi)}} + \minn{u'}{\minn{f \in \tilde{\C}_{\thj+1}}{ f(u', \phi)} }+ \lambda\\
&= \minn{f \in \C_{\thj}}{\minn{u}{ f(u, \phi)}} + \minn{f \in \tilde{\C}_{\thj+1}}{\minn{u'}{ f(u', \phi)} }+ \lambda. \label{eq:c-phi-thj-ch-univariate}
\end{align}
Since $f(u, \phi)$ is piecewise quadratic in $u$ and $\phi$ (Theorem~\ref{thm:2d-pruning}), we see that  $\minn{u}{ f(u,\phi)}$ is piecewise quadratic in $\phi$. 
Therefore,   $ \minn{f \in \C_{\thj}}{\minn{u}{ f(u, \phi)}}$ and $ \minn{f \in \tilde\C_{\thj+1}}{\minn{u}{ f(u, \phi)}}$ can be efficiently performed using ideas from \cite{rigaill2015pruned} and \cite{maidstone2017optimal}, which 
allow for efficient manipulations of piecewise quadratic functions of a single variable. This means that $C_{\thj}'(\phi)$ can be efficiently computed. Recall from Theorem~\ref{thm:2d-pruning} that the set $\C_{\thj}$ contains $\thj - \thi{j-1} + 1$ functions and can be computed in $\mathcal{O}((\thj - \thi{j-1})^{2})$ operations. Therefore, computing $C_{\thj}'(\phi)$ requires $\mathcal{O}((\thj - \thi{j-1})^{2})$ operations to compute $\C_{\thj}$
and $\mathcal{O}((\thi{j+1} - \thi{j})^{2})$ operations to compute $\tilde\C_{\thj+1}$, followed by performing the operation $\min_{u}\{f(u, \phi)\}$ a total of $\mathcal{O}(\thi{j+1}- \thi{j-1})$ times. We can similarly obtain the piecewise quadratic function $C_{\neg\thj}'(\phi)$ of $\phi$. Therefore, we can analytically compute $C'(\phi)$.  

Finally, recall from \eqref{eq:set-S2} that  $\mathcal{S}=\{\phi: C(\phi) \leq C'(\phi)\}$. Since we have efficiently characterized both $C(\phi)$ and $C'(\phi)$, our characterization of $\mathcal{S}$ is complete. %, the ar

\section{Experiments}
\label{sec:experiments}

\subsection{Simulation set-up and methods for comparison} 
\label{sec:exp-intro}

\begin{figure}[h!]
\centering
\includegraphics[width = 0.8\textwidth]{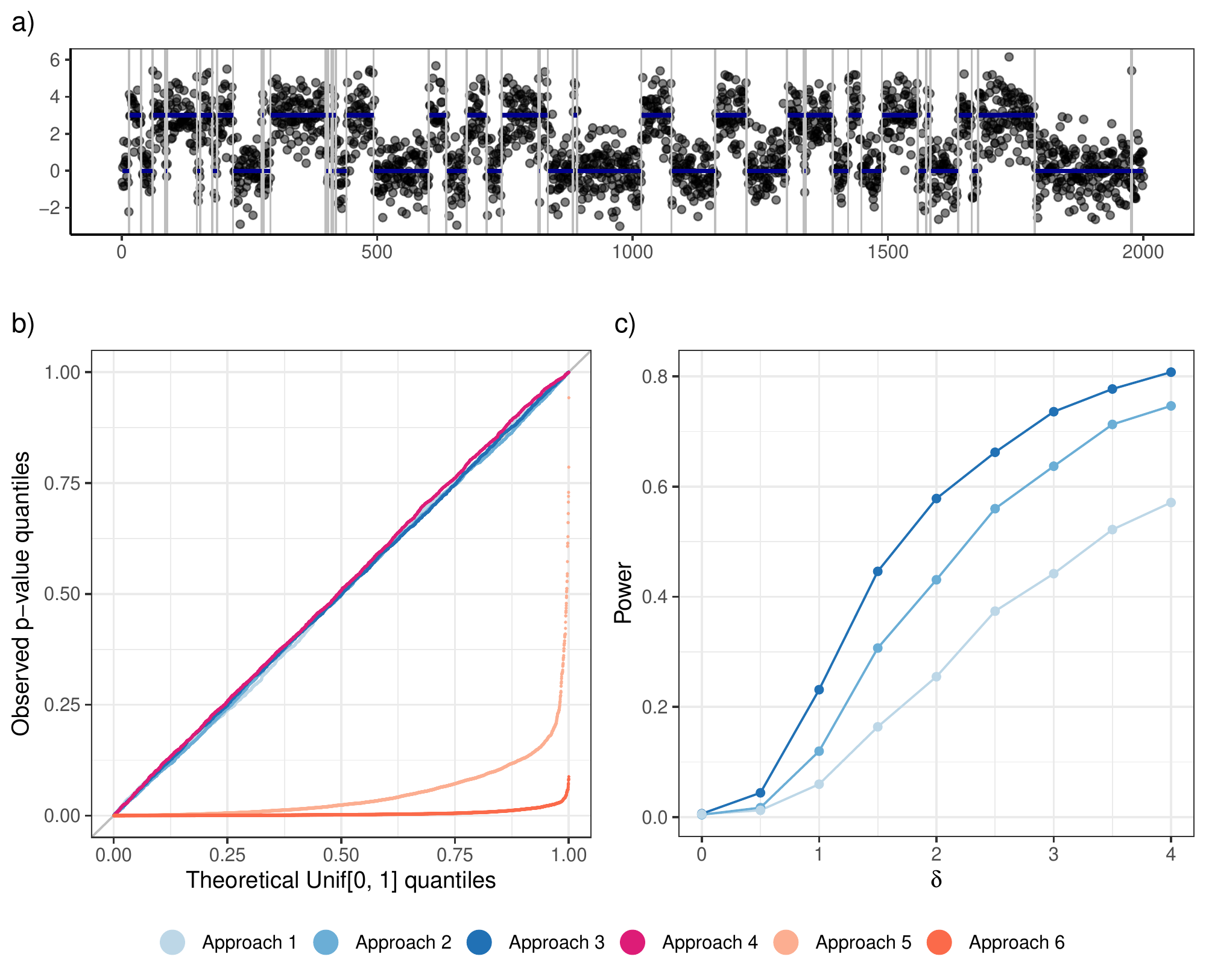}
\caption{a) The grey points represent a realization from the mean model \eqref{eq:obs-model}, with true change in mean due to a changepoint $\delta=3$. The mean $\mu_1,\ldots,\mu_T$ is shown as a blue line, and the changepoints are shown as grey vertical lines. 
b) Quantile-quantile plot comparing sample $p$-value quantiles under \eqref{eq:obs-model} with $\mu_1=\ldots=\mu_{2000}$ versus theoretical quantiles of the $\mathrm{Unif}(0, 1)$ distribution, for Approaches 1--4 in Section~\ref{sec:exp-intro}, and Approaches 5--6 in Section~\ref{sec:T1E}. c) 
 Empirical power, averaged over 100 replicates, is displayed for Approaches 1--3 defined in Section~\ref{sec:exp-intro}, each of which results from testing $H_0: \nu^{\top} \mu=0$ for changepoints estimated using binary segmentation with different conditioning sets. 
 Various values of  $\delta$, the true change in mean due to a changepoint, are shown on the $x$-axis. 
 Power increases with the size of the conditioning set.
  }
\label{fig:bs_delta_diff_seg_lengths}
\end{figure}

We simulate $y_{1}, \ldots, y_{2000}$ according to \eqref{eq:obs-model} with $\sigma^{2} = 1$.
The mean vector $\mu \in \mathbb{R}^{2000}$ is piecewise constant with $50$ changepoints. After each even-numbered changepoint the mean equals $0$, and after each odd-numbered changepoint it equals $\delta$, 
for 
%when the ; between consecutive changepoints, it equals either $0$ or $\delta$, for 
%for the $\tau_j$th and $(\tau_{j+1})$st timepoint, either $\mu_{\tau_j+1}=\ldots=\mu_{\tau_{j+1}}=0$ or $\mu_{\tau_j+1}=\ldots=\mu_{\tau_{j+1}}=\delta$
%, for 
%
% and with $\mu$ defined as 
%$\mu_{\tau_{j}+1} = \mu_{\tau_{j}+2} = \cdots = \mu_{\tau_{j+1}} := \delta 1_{\{j \mbox{ odd}\}}$ for 
 $\delta \in \{ 0, 0.5$, $1.0$, $1.5$, $2.0$, $2.5$, $3.0$, $3.5$, $4.0\}$. The $K = 50$ changepoints are sampled without replacement from $\{1, 2, \ldots, 1999\}$. Panel a) of Figure~\ref{fig:bs_delta_diff_seg_lengths} depicts a  realization with $\delta = 3$. 

We compare four tests of a change in mean at an estimated changepoint: 
\begin{list}{}{}
\item{\emph{Approach 1.}} For the $j$th changepoint estimated by binary segmentation, test $H_0: \nu^{\top} \mu=0$ using  $\nu$ in \eqref{eq:nu}. Condition on the locations, orders, and signs of all of the estimated changepoints from binary segmentation. This is closely related to \cite{hyun2018post}'s proposal. 
\vspace{-2mm}
\item{\emph{Approach 2.}} For the $j$th changepoint estimated by binary segmentation, test $H_0: \nu^{\top} \mu=0$ using  $\nu$ in \eqref{eq:nu}. Condition on the locations of all of the estimated  changepoints from binary segmentation. %\textcolor{green}{This was discussed in Section~\ref{??}.}
\vspace{-2mm}
\item{\emph{Approach 3.}} For the $j$th changepoint estimated by binary segmentation, test $H_0: \nu^{\top} \mu=0$ using  $\nu$ in \eqref{eq:nu-window}. Condition only on the location of the $j$th estimated changepoint from binary segmentation.  %\textcolor{green}{This was discussed in Section~\ref{??}.}
\vspace{-2mm}
\item{\emph{Approach 4.}} For the $j$th changepoint estimated by $\ell_0$ segmentation, test $H_0: \nu^{\top} \mu=0$ using  $\nu$ in \eqref{eq:nu-window}. Condition only on the location of the $j$th estimated changepoint from $\ell_0$ segmentation.  %\textcolor{green}{This was discussed in Section~\ref{??}.}
\end{list} 
 Unless stated otherwise, we  take $h=50$ in \eqref{eq:nu-window} for Approaches 3--4.  As our aim is to compare the power of Approaches 1--4, we assume the true number of changepoints ($K=50$) is known,  so that both binary segmentation and $\ell_{0}$ segmentation estimate the same number of changepoints\footnote{On a given data set, there may not exist a  $\lambda$ such that $\ell_0$ segmentation yields precisely $50$ estimated changepoints. In this case, we select $\lambda$ to give approximately $50$ estimated changepoints.}. We also assume that the underlying noise variance ($\sigma^{2} = 1$) is known; see Section~\ref{sec:extensions} for a more detailed discussion. All results  are averaged over $100$ replicate data sets with $\mu$ fixed.  
 
% In Approaches~1--3, we approximate the set $\mathcal{S}$ with $\tilde{\mathcal{S}}$ as described in Proposition~\ref{prop:early-stopping}; we take $|a_{-r}| = |a_{r'+1}| = \max( 10 \sigma||\nu||_{2}, |\nu^{\top}y| )$. 
 
In Section~\ref{sec:timing-relative} of the Supplementary Materials, we present timing results for estimating changepoints as well as computing $p$-values using Approaches 1--4. Surprisingly, Approach~4 is even faster than Approaches 1--3: in our C++ implementation, the former takes only 15 seconds when $T=1000$. Approaches 1--3 take longer because calculating $\mathcal{S}$ in the case of binary segmentation requires manipulating a large set of linear equations.

\subsection{Type I error control under a global null} \label{sec:T1E}

We take $\delta=0$, so that $\mu_1=\ldots=\mu_{2000}$, and consider testing $H_0: \nu^{\top} \mu=0$ using Approaches 1--4, as well as  the following two approaches that rely on a standard $z$-test:
\begin{list}{}{}
\item{\emph{Approach 5.}} For the $j$th changepoint estimated by binary segmentation, test $H_0: \nu^{\top} \mu=0$ using  $\nu$ in \eqref{eq:nu}, without conditioning.
% for the changepoints estimated by binary segmentation, using a  standard \textcolor{green}{$z$-test}.
\vspace{-3mm}
\item{\emph{Approach 6.}}  For the $j$th changepoint estimated by $\ell_0$ segmentation, test $H_0: \nu^{\top} \mu=0$ using  $\nu$ in \eqref{eq:nu}, without conditioning. 
 %for the changepoints estimated by $\ell_0$ segmentation, using a  standard \textcolor{green}{$z$-test}.
\end{list} 
These two  approaches do not account for the fact that the changepoints were estimated from the data.  Panel b) of Figure~\ref{fig:bs_delta_diff_seg_lengths}  displays quantile-quantile plots of the observed $p$-value quantiles versus theoretical $\mathrm{Unif}[0,1]$ quantiles.  
 The plots indicate that Approaches~1--4 control the Type 1 error, whereas Approaches~5--6 do not.

\subsection{Increases in power due to conditioning on less information} 
\label{sec:increases-power-smaller-sets}

Next, we illustrate that the power increases as the size of the conditioning set increases, by considering Approaches 1--3 from Section~\ref{sec:exp-intro}. Each approach uses binary segmentation,  though with different   conditioning sets. 

On a given dataset, we define the empirical power as the fraction of true changepoints for which the nearest estimated changepoint has a $p$-value below $\alpha$ and is within $\pm m$ timepoints, 
\begin{align}
\widehat{\text{Power}} := \frac{ \sum_{i = 1}^{K} 1_{\left( {|\tau_{i} - \thi{j(i)}|}\leq m \text{ and }  p_{j(i)} \leq \alpha	\right)}}{K}.
\label{eq:est-power}
\end{align}
Here,  $j(i) = \mathrm{argmin}_{1 \leq l \leq K }{|\tau_{i} - \thi{l}|}$. 
Panel c) of Figure~\ref{fig:bs_delta_diff_seg_lengths} shows the empirical power 
for the three approaches with $\alpha = 0.05$ and $m = 2$. As the size of the conditioning set increases, from $\{\phi : \mathcal{M}(y'(\phi)) = \mathcal{M}(y), \mathcal{O}(y'(\phi)) = \mathcal{O}(y), \Delta(y'(\phi)) = \Delta(y) \}$ to 
$\{\phi : \mathcal{M}(y'(\phi)) = \mathcal{M}(y)\}$ to $\{\phi : \thj \in \mathcal{M}(y'(\phi)) \}$, the power increases substantially.

\subsection{Power and detection probability} 
 \label{sec:power-detection-prob}
 
We now compare the performances of Approaches 1--4, defined in Section~\ref{sec:exp-intro}, as well as two additional approaches that are based on \emph{sample splitting} \citep{cox1975note}: 
 \begin{list}{}{}
\item{\emph{Approach 7.}}  Apply binary segmentation to the odd timepoints. For the $j$th estimated changepoint, test  $H_0: \nu^{\top} \mu=0$ on the even timepoints, with $\nu$ in \eqref{eq:nu}, without conditioning. 
\vspace{-3mm}
\item{\emph{Approach 8.}} Apply $\ell_0$ segmentation to the odd timepoints. For the $j$th estimated changepoint, test $H_0: \nu^{\top} \mu=0$ on the even timepoints, with $\nu$ in \eqref{eq:nu}, without conditioning. 
\end{list}
 Because we estimate and test the changepoints on two separate halves of the data,  we can apply a standard $z$-test in Approaches 7 and 8 \citep{fithian2014optimal}. 

In addition to calculating the empirical power  \eqref{eq:est-power} for each approach, we also consider each approach's ability to detect the true changepoints. This is defined as the fraction of true changepoints for which there is an estimated changepoint within $\pm m$ timepoints, 
  \begin{equation}
 \widehat{\text{Detection probability}} := \frac{ \sum_{i = 1}^{K} 1_{\left( \min_{1 \leq l \leq K }{|\tau_{i} - \thi{l} | }\leq m \right)}}{K}.
 \label{eq:detection-prob}
 \end{equation}

\begin{figure}[h!]
\includegraphics[width = \textwidth]{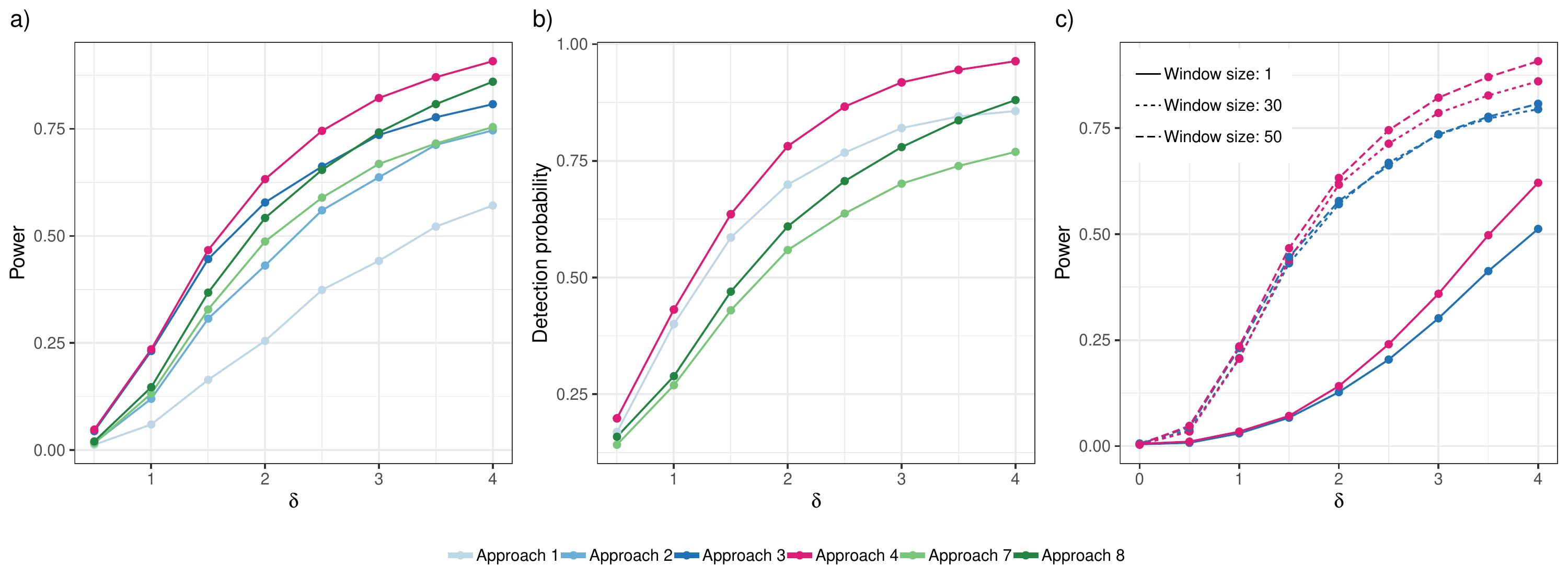}
\caption{Empirical power and detection probability for different changepoint estimation and inference procedures. a) Power for Approaches 1--4, which are  described in Section~\ref{sec:exp-intro}, as well as Approaches 7--8, which are  described in Section~\ref{sec:power-detection-prob}. b) Detection probability for binary segmentation and $\ell_0$ segmentation using all of the data, as well as half of the data. In this panel, the curve shown for Approach~1 applies to Approaches~1-3 since Approaches~1-3 use binary segmentation. c) Power of Approaches 3 and 4  for testing $H_0: \nu^\top \mu=0$ for $\nu$ in \eqref{eq:nu-window}, for three values of the window size $h$.}
\label{fig:l0_bs_delta_diff_seg_lengths}
\end{figure}

Panels b) and c) of Figure~\ref{fig:l0_bs_delta_diff_seg_lengths} display the power and detection probability for Approaches 1--4 and 7--8, with $\alpha = 0.05$ and $m = 2$. Approach 4 (which makes use of $\ell_0$ segmentation, and conditions only on the $j$th estimated changepoint) performs the best, in terms of both power and detection probability, especially as $\delta$ increases. 
% yields the highest power, especially for larger values of $\delta$; it has the highest detection probability. 
%. In panel b),  $\ell_{0}$ segmentation vastly outperforms binary segmentation in terms of its ability to detect true changepoints.  
  Figure~\ref{fig:l0_bs_delta_diff_seg_lengths} also illustrates the benefit of the inferential framework developed in this paper over naive sample-splitting approaches. Sample splitting has limited ability to detect changepoints, since only half of the data is used to estimate changepoints. 

\subsection{Assessment of different window sizes for testing $H_0: \nu^\top \mu=0$ for $\nu$ in \eqref{eq:nu-window}}

% Recall that Approach 4 tests $H_0: \nu^\top \mu$ for $\nu$ in \eqref{eq:nu-window}, and conditions on  just $\thj \in \mathcal{M}(y'(\phi))$. 
 Figure~\ref{fig:l0_bs_delta_diff_seg_lengths} suggests that Approaches 3 and 4 from Section~\ref{sec:exp-intro}   have high power. However, they require pre-specifying the window size $h$ in \eqref{eq:nu-window}. We now address this possible weakness. In Figure~\ref{fig:l0_bs_delta_diff_seg_lengths}c), we assess the performance of Approaches 3 and 4 with  $h \in \{1, 30, 50 \}$. Provided that $h$ is sufficiently large, its value has little effect on the power.

\section{Real data example} 
\label{sec:real-data-example}

We now consider guanine-cytosine (G-C) content on a 2Mb window of human chromosome one, binned so that $T = 2000$. Data was originally accessed from the National Center for Biotechnology Information, and is available via the \texttt{R} package \texttt{changepoint} \citep{killick2014changepoint}. We used a consistent estimator of $\sigma$ described in Section~\ref{sec:extensions} to scale the data and calculate $p$-values.

\begin{figure}[h!]
\includegraphics[width = \textwidth]{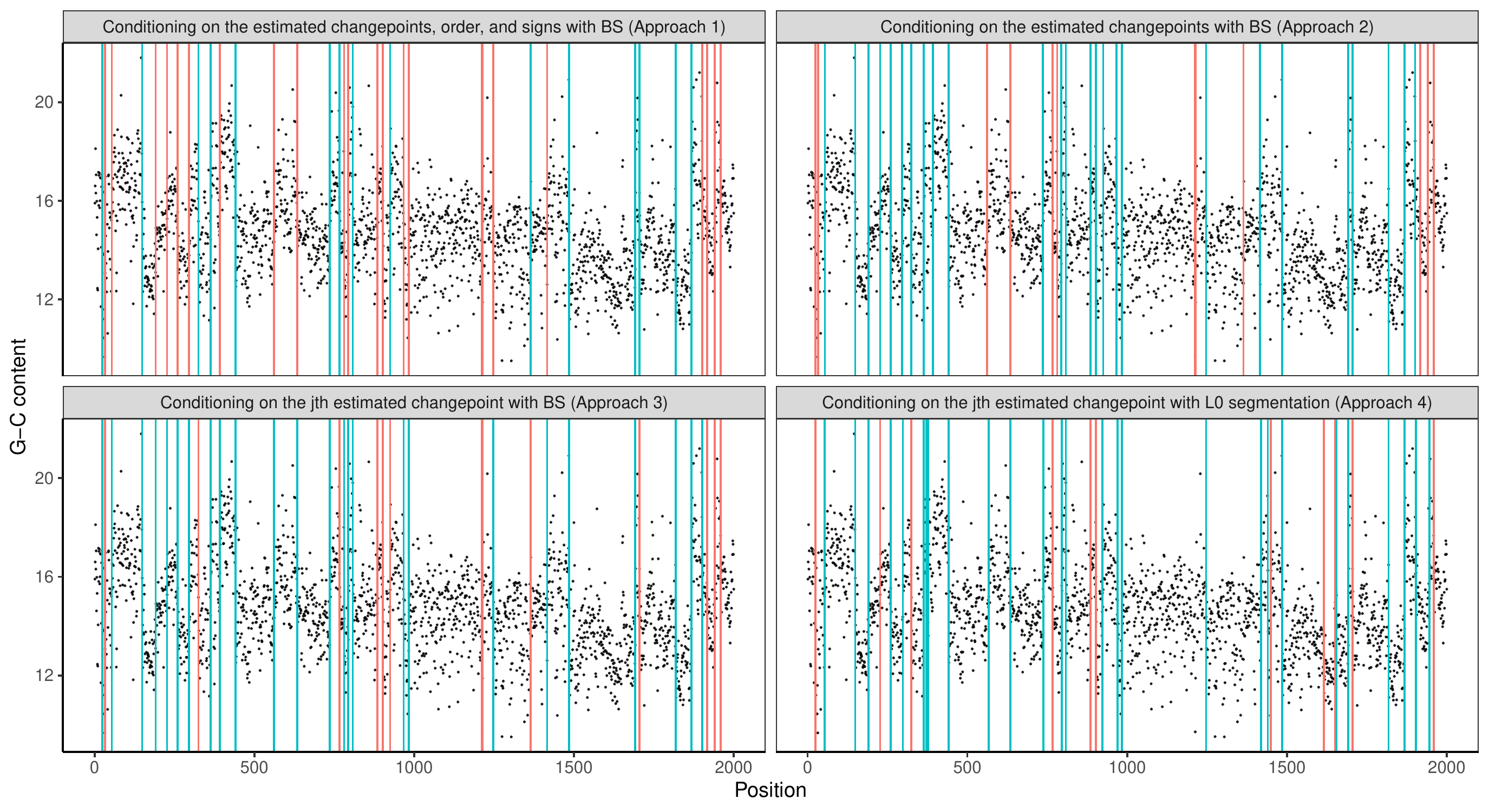}
\caption{The number of discoveries depends on the size of the conditioning set. 
Each panel displays scaled G-C content on a 2Mb window of human chromosome one. The G-C content is binned leading to $T = 2000$ (displayed in black). 
Estimated changepoints from Approaches 1--4 from Section~\ref{sec:exp-intro} (organized by panel) for which the $p$-value is less than $0.05$ are displayed in blue; the remaining estimated changepoints are displayed in red.}
\label{fig:real-data}
\end{figure}

We estimate changepoints using $k$-step binary segmentation, where $k=38$ is chosen based on the modified Bayesian information criterion \citep{zhang2007modified} implemented in the \texttt{changepoint} package. To facilitate comparisons, we then fit $\ell_{0}$ segmentation using a value of $\lambda$ that yields $38$ changepoints. Figure~\ref{fig:real-data} displays the estimated changepoints from these two methods, along with an indication of whether Approaches 1--4 from Section~\ref{sec:exp-intro} resulted in a $p$-value below $0.05$. The number of discoveries (estimated changepoints whose $p$-value is less than $0.05$) is substantially greater using Approaches 2--4 than using Approach 1, which conditions on far more information. Approach~1 results in 15 discoveries, versus 26, 25, and 27  in Approaches 2, 3, and 4, respectively.  
These $p$-values can be adjusted for multiple testing using ideas from e.g. \cite{benjamini1995controlling}, \cite{storey2002direct}, and \cite{dudoit2007multiple}.

\section{Discussion}
\label{sec:discussion}

% that testing for a change in mean around an estimated changepoint simply requires characterizing the set $\mathcal{S}$, defined in either \eqref{eq:set-s} or \eqref{eq:set-s-thj}. We  introduce the necessary computational tools to do this for three popular changepoint detection algorithms. Importantly, since our approach does not rely on 
%the polyhedral lemma of \cite{lee2016exact},
%the conditioning sets that we use are much larger than those  in earlier work and lead to higher-powered tests. 
% We now discuss a few extensions of our work. 

\subsection{Larger conditioning sets for testing \eqref{eq:p2} with $\nu$ in \eqref{eq:nu}}
\label{sec:smaller-conditioning-p2}

No special properties of the conditioning set were used to prove Theorem~\ref{thm:pval}. Thus, instead of conditioning on the full set of changepoints as  in Section~\ref{sec:biggerconditioning}, we could have instead conditioned on the $j$th estimated changepoint and its immediate neighbors. This would yield the $p$-value 
$p = \text{Pr}\left(  | \phi | \geq | \nu^{\top} y | \mid \{\thi{j-1}, \thj, \thi{j+1}\} \subseteq \mathcal{M}(y'({\phi}))  \right)$.
Characterizing the set $\mathcal{S}=\{ \phi: \{\thi{j-1}, \thj, \thi{j+1}\} \subseteq \mathcal{M}(y'({\phi})) \}$ would require only
minor modifications to the algorithms in Sections~\ref{sec:bs-characterization} and \ref{sec:l0-characterization} and the Supplementary Materials. %Based on our simulation results, we suspect that this would yield increased power relative tot % and in the Supplemental Materials. 

%For some conditioning sets and changepoint detection algorithms, it might be difficult to characterize $\mathcal{S}$. In this case, it is still possible to approximate $\mathcal{S}$ by testing whether or not $\phi\in \mathcal{S}$ for a fine grid of $\phi$ values; this approach is also suggested by \cite{liu2018more}. 

%\subsection{Confidence intervals for the change in mean}

%To construct confidence intervals for the change in mean, we first define $H_{0}(c): \nu^{T}\mu = c$. We note that since
%$
%\mathbb{C}(\phi) =  \left\{c \text{ } | \text{ } \text{Pr}_{H_{0}(c)} \left( |\phi| \geq |\nu^{\top}y| \mid \phi \in \mathcal{S}\right) \geq \alpha \right\}$ satisfies 
%$
%\text{Pr}(\nu^{\top}\mu \in \mathbb{C}(\phi) | \phi \in \mathcal{S}) \geq 1 - \alpha,
%$
%the set $\mathbb{C}(\phi)$ is a $100(1- \alpha)\%$ confidence interval for $\nu^{\top}\mu$. Importantly, we can efficiently calculate $\mathbb{C}(\phi)$ since the set $\mathcal{S}$ is unchanged as we vary $c$; only the mean of the null distribution for $\nu^{\top}Y$ changes. 

\subsection{Extensions to related problems} 

The ideas in this paper apply beyond the change-in-mean model \eqref{eq:obs-model}. 
%In particular, the ideas in Section~\ref{sec:selective-inference} only require conditioning on the sufficient statistics of $\nu^{\top}Y$. 
%
% 
For instance, they can be applied to the analysis of data from calcium imaging, a recent technology for recording neuronal activity \emph{in vivo} \citep{dombeck2007imaging}. A number of authors \citep{vogelstein2010fast, friedrich2017fast} have assumed that the observed fluorescence trace for a neuron, $y_{t}$, is a noisy version of the underlying calcium concentration, $c_{t}$,  which decays exponentially with a rate $\gamma<1$, except when there is an instantaneous increase in the calcium because the neuron has spiked, $s_{t} > 0$:
\begin{align*}
Y_{t} &= c_{t} + \epsilon_{t}, \quad \epsilon_{t} \overset{\text{iid}}{\sim} N(0, \sigma^{2}),  \quad
c_{t} = \gamma c_{t-1} + s_{t}.
\end{align*}
In this model, scientific interest lies in determining the precise timepoints of the spikes, i.e. the set $\{t: s_t > 0\}$. \cite{jewell2018exact} and \cite{jewell2018fast} estimate this quantity by solving 
a variant of the $\ell_{0}$ segmentation problem \eqref{eq:l0-mean-rss} in Section~\ref{sec:l0-opt-chgpt}. The framework from Section~\ref{sec:selective-inference}, and the algorithms from Section~\ref{sec:l0-characterization}, can be used to test the null hypothesis that there is no increase in the calcium concentration around a spike, $H_{0}:\nu^{\top}c=0$, for a suitably chosen contrast $\nu$.  
%Furthermore, the algorithms developed in Section~\ref{sec:l0-characterization} can be modified to efficiently characterize the selective distribution; 
Details are  in \cite{chen2021quantifying}.

It is natural to wonder whether these ideas can be extended to  the change-in-slope proposals of \cite{fearnhead2018detecting} and \cite{baranowski2019narrowest}. Extending the ideas in Section~\ref{sec:l0-characterization} to the former is quite challenging, since the continuity constraint in the optimization problem induces dependence across segments that complicate the development of computationally-feasible recursions. By contrast, the latter is closely related to binary segmentation, and so an extension of the approach in Section~\ref{sec:bs-characterization}  can be applied.

\subsection{Additional extensions}
\label{sec:extensions}

\paragraph{Relaxing assumptions in \eqref{eq:obs-model}}
The model \eqref{eq:obs-model}   assumes that the error terms are Gaussian, independent, and identically distributed. 
These assumptions are critical to the proof of Theorem~\ref{thm:pval}, as they guarantee that $\nu^\top Y$ and $\Pi_\nu^\perp Y$ are independent. However, recent work in selective inference has focused on relaxing these assumptions \citep{tian2018selective,tibshirani2018uniform,taylor2018post}, and may be applicable here.

\paragraph{Estimation of the error variance in \eqref{eq:obs-model}}

Throughout this paper, we have assumed that the error variance in \eqref{eq:obs-model} is known. However, if it is unknown, then we can plug in any consistent estimator of $\sigma$ in evaluating the $p$-values in \eqref{eq:p2} and \eqref{eq:p2-thj}. Then, under $H_0: \nu^\top \mu=0$, the resulting $p$-values will converge in distribution to a $\mathrm{Unif}[0,1]$ distribution, i.e. they will have asymptotic Type 1 error control. In Section~\ref{sec:performance-unknown-sigma} of the Supplementary Materials, we present the results of a simulation study using a simple consistent estimator of $\sigma$ obtained by taking the median absolute deviation of the first differences of  $y_1,\ldots,y_T$ and scaling by $\sqrt{2} \Phi^{-1} (3/4)$. We see that this approach leads to adequate Type 1 error control, as well as substantial power under the alternative.

\paragraph{Confidence intervals}
 The conditional distribution of $\nu^\top Y$ can be used to develop a confidence interval for  $\nu^\top \mu$ that has correct selective coverage; see, e.g., \cite{lee2016exact}.

\section*{Acknowledgments}
Sean Jewell received funding from the Natural Sciences and Engineering Research Council of Canada. This work was partially supported by Engineering and Physical Sciences Research Council Grant EP/N031938/1 to Paul Fearnhead, and NSF CAREER DMS-1252624, NIH grants DP5OD009145, R01DA047869, and R01EB026908, and a Simons Investigator Award in Mathematical Modeling of Living Systems to Daniela Witten.

We thank Zaid Harchaoui and Ali Shojaie for helpful conversations, and Jacob Bien and three anonymous reviewers for suggestions that improved the quality of this paper.

\bibliographystyle{apalike}
\bibliography{bib}

\makeatletter\@input{xx.tex}\makeatother

\end{document}

\maketitle
\appendix

\section{Proof of Theorem~\ref{thm:pval}}
\label{sec:proof-pval}

To characterize \eqref{eq:p2}, we note that $Y$ decomposes as
\begin{align}
Y = (I - \Pi_{\nu}^{\perp})Y + \Pi_{\nu}^{\perp}Y,
\label{eq:y-decompose}
\end{align}
where $\Pi_{\nu}^{\perp} = I - \frac{\nu\nu^{\top}}{||\nu||_{2}^{2}}$.
Then \eqref{eq:p2} becomes 
\begin{align}
p  & = \text{Pr}_{H_0} \left(  | \nu^{\top} Y |  \geq | \nu^{\top} y | \mid \mathcal{M}(Y) = \mathcal{M}(y)  , \Pi_\nu^{\perp} Y = \Pi_\nu^{\perp} y \right) \label{eq:sub}\\
%&= \text{Pr}_{H_0} \left(  | \nu^{\top} Y |  \geq | \nu^{\top} y | \mid \mathcal{M}(Y) = \mathcal{M}(y)  , \Pi_\nu^{\perp} Y = \Pi_\nu^{\perp} y, Y = (I - \Pi_{\nu}^{\perp})Y + \Pi_{\nu}^{\perp}y \right) \\
&= \text{Pr}_{H_0} \left(  | \nu^{\top} Y |  \geq | \nu^{\top} y | \mid \mathcal{M}((I - \Pi_{\nu}^{\perp})Y + \Pi_{\nu}^{\perp}y ) = \mathcal{M}(y)  , \Pi_\nu^{\perp} Y = \Pi_\nu^{\perp} y \right) \label{eq:prop-orth} \\
%&= \text{Pr}_{H_0} \left(  | \nu^{\top} Y |  \geq | \nu^{\top} y | \mid \mathcal{M}((I - \Pi_{\nu}^{\perp})Y + \Pi_{\nu}^{\perp}y ) = \mathcal{M}(y)  , \Pi_\nu^{\perp} Y = \Pi_\nu^{\perp} y\right) \\
&= \text{Pr}_{H_0} \left(  | \nu^{\top} Y |  \geq | \nu^{\top} y | \mid \mathcal{M}((I - \Pi_{\nu}^{\perp})Y + \Pi_{\nu}^{\perp}y ) = \mathcal{M}(y) \right).  \label{eq:independence}
\end{align}
Here, \eqref{eq:sub} is our definition of a $p$-value \eqref{eq:p2}, and \eqref{eq:prop-orth} follows from \eqref{eq:y-decompose} and the fact that $\Pi_\nu^{\perp} Y = \Pi_\nu^{\perp} y$. Finally, \eqref{eq:independence} follows from the fact that $Y$ is Gaussian (see \eqref{eq:obs-model}) and so  $\nu^{\top} Y$ and $\Pi_\nu^\perp Y$ are independent. 

Moreover, we note that \eqref{eq:obs-model} implies that $\nu^{\top} Y \sim N(\nu^{\top} \mu, \| \nu \|^2 \sigma^2)$, and 
that under the null hypothesis \eqref{eq:hyp-test}, $\nu^{\top} Y \sim N(0, \| \nu \|^2 \sigma^2)$. We now define $\phi = \nu^{\top}Y$; thus under the null hypothesis, $\phi \sim N(0, \| \nu \|^2 \sigma^2)$. Recall that
\begin{align}
y'(\phi)= y - \frac{\nu \nu^{\top}y}{||\nu||_{2}^{2}}  + \frac{\nu \phi}{||\nu||_{2}^{2}}.
\label{eq:yphi-proof}
\end{align}
Therefore, 
 \begin{equation}
 p = \text{Pr}\left(  | \phi | \geq | \nu^{\top} y | \mid \mathcal{M}(y'({\phi})) = \mathcal{M}(y)   \right).
 \label{eq:pfinal-proof}
 \end{equation}

\section{Details related to Section~\ref{sec:bs-characterization}}
\label{sec:efficient-bs-supp}

\subsection{Proof of Proposition~\ref{prop:s-decomposition}}
\label{sec:proof-prop-s-decomposition}

 To prove the first statement in Proposition~\ref{prop:s-decomposition}, we note from Proposition~\ref{prop:hyun} that the set of data that yields changepoints $m$, orders $o$, and signs $d$ is of the form $\{ y : \bm{\Gamma} y \leq 0 \}$. Therefore, the set of $\phi$ that yields $\mathcal{M}(y'(\phi)) = m, \mathcal{O}(y'(\phi)) = o,$ and  $\Delta(y'(\phi)) = d$ is of the form
$\{ \phi : \bm{\Gamma} y'(\phi) \leq 0 \}$.
Since $\bm{\Gamma} y'(\phi) \leq 0$ represents $k(2T - k - 3)$ linear inequalities in $\phi$, the set $\{ \phi : \bm{\Gamma} y'(\phi) \leq 0 \}$ is an interval. 

The second statement in Proposition~\ref{prop:s-decomposition} follows from the fact that
\begin{align}
\mathcal{S} &= \bigcup_{o \in O, d \in D} \{\phi : \mathcal{M}(y'(\phi)) = \mathcal{M}(y), \mathcal{O}(y'(\phi)) = o, \Delta(y'(\phi)) = d \} \label{eq:basic-enumeration} \\\
&= \bigcup_{(o', d') \in \mathcal{I}} \{\phi : \mathcal{M}(y'(\phi)) = \mathcal{M}(y), \mathcal{O}(y'(\phi)) = o', \Delta(y'(\phi)) = d' \} \label{eq:in-J} \\
&= \bigcup_{i \in \mathcal{J}} [a_{i}, a_{i+1}] \label{eq:interval}
\end{align}
where 
$O$ is the set of cardinality $k!$ containing all possible orders of the $k$ changepoints, and
$D := \{-1, + 1\}^{k}$ is the set of possible signs.  \eqref{eq:basic-enumeration} follows from the definition of $\mathcal{S}$. 
\eqref{eq:in-J} follows from the definition of $\mathcal{I}$ in \eqref{eq:num-elements-i}. 
 \eqref{eq:interval} results from applying Proposition~\ref{prop:hyun} to each set of the form $\{\phi : \mathcal{M}(y'(\phi)) = \mathcal{M}(y), \mathcal{O}(y'(\phi)) = o', \Delta(y'(\phi)) = d' \}$. 
 Therefore, $|\mathcal{J}| = |\mathcal{I}|$.

%\eqref{eq:interval} follows from \eqref{eq:in-J} since for each $(o', d')$ in $\mathcal{J}$, Proposition~\ref{prop:hyun} can be used to obtain an interval $(a, b) := \{\phi : \bm{\Gamma}(o', d') y'(\phi) \leq 0 \}$ such that $\mathcal{O}(y'(\phi)) = o'$, $\Delta(y'(\phi)) = d'$, and $\mathcal{M}(y'(\phi)) = \mathcal{M}(y'(\alpha))$ for the $\alpha$ corresponding to $(o', d')$. Here, to prevent confusion, the matrix $\bm{\Gamma}$ is written as a function of the order and signs of the changepoints, $\bm{\Gamma}(o', d')$. If $\mathcal{M}(y'(\alpha)) = \mathcal{M}(y)$ then $[a, b]\in\mathcal{S}$. \hfill $\square$

It is worth noting that \eqref{eq:basic-enumeration} is the union over $2^{k} k!$ intervals. By contrast, \eqref{eq:interval} is a union over $|\mathcal{J}| = |\mathcal{I}|$ intervals, which in practice is much smaller than $2^{k} k!$.

\subsection{Proof of Proposition~\ref{prop:early-stopping}}
\label{sec:reducing-comp-cost-bs}

To prove Proposition~\ref{prop:early-stopping}, recall that $\mathcal{S} = \bigcup_{i \in \mathcal{J}} [a_{i}, a_{i+1}]$, as described in Section~\ref{sec:bs-characterization}, where there exist $i, i'\in \mathcal{J}$ such that $a_{i} = -\infty$ and $a_{i' + 1} = \infty$. 
Also recall that $\tilde{\mathcal{S}} = (-\infty, a_{-r}] \cup \left( \bigcup_{i \in \mathcal{J}\cap \{-r, \ldots, r'\} } [a_{i}, a_{i+1}]\right) \cup [a_{r' + 1}, \infty)$, for some $r$ and $r'$ such that $a_{-r} \leq -|\nu^{\top}y|$ and $a_{r'+1} \geq |\nu^{\top}y|$. Since $\{ \phi: |\phi | \geq |\nu^{\top}y|\} \cap \{ \phi: \phi \in \tilde{\mathcal{S}} \setminus \mathcal{S}\} = \{ \phi: \phi \in \tilde{\mathcal{S}} \setminus \mathcal{S}\}$, we have that
\begin{align*}
\text{Pr}(|\phi| \geq |\nu^{\top}y | \mid \phi \in \tilde{\mathcal{S}}) &= \frac{\text{Pr}(\{ |\phi | \geq |\nu^{\top}y|\} \cap \{ \phi \in \tilde{\mathcal{S}}\})}{\text{Pr}(\phi \in \tilde{\mathcal{S}})} \\
&= \frac{\text{Pr}(\{ |\phi | \geq |\nu^{\top}y|\} \cap \{ \phi \in \mathcal{S}\}) + 
\text{Pr}(\{ |\phi | \geq |\nu^{\top}y|\} \cap \{ \phi \in \tilde{\mathcal{S}} \setminus \mathcal{S}\})
}{\text{Pr}(\phi \in \mathcal{S}) + \text{Pr}(\phi \in \tilde{\mathcal{S}} \setminus \mathcal{S})} \\
&= \frac{\text{Pr}(\{ |\phi | \geq |\nu^{\top}y|\} \cap \{ \phi \in \mathcal{S}\}) + 
\text{Pr}( \phi \in \tilde{\mathcal{S}} \setminus \mathcal{S})
}{\text{Pr}(\phi \in \mathcal{S}) + \text{Pr}(\phi \in \tilde{\mathcal{S}} \setminus \mathcal{S})} \\
&\geq \text{Pr}(|\phi| \geq |\nu^{\top}y | \mid \phi \in \mathcal{S}).
\end{align*}

\subsection{Characterization of \eqref{eq:set-s-thj}}
\label{sec:basic-set-s-character-bs-2}

In this section, we show that we can characterize the set $\mathcal{S} \equiv \{ \phi: \thj \in \mathcal{M}(y'(\phi)) \}$ for changepoints estimated via binary segmentation. 
Our approach is very similar to that of Section~\ref{sec:bs-characterization}. In the following two propositions, Propositions~\ref{prop:s-decomposition-thj} and \ref{prop:efficient-intervals-thj}, we modify Propositions~\ref{prop:s-decomposition}
and \ref{prop:efficient-intervals} for the case of $\mathcal{S}$ defined in \eqref{eq:set-s-thj}. 

\begin{prop}
\label{prop:s-decomposition-thj}
The set $\{\phi : \mathcal{M}(y'(\phi)) = m, \mathcal{O}(y'(\phi)) = o, \Delta(y'(\phi)) = d \}$ is an interval.
Furthermore, the set $\mathcal{S}$ defined in \eqref{eq:set-s-thj} can be written as the union of intervals,
\begin{align}
\mathcal{S} = \{ \phi: \thj \in \mathcal{M}(y'(\phi)) \} &= \bigcup_{i \in \mathcal{J}} [a_{i}, a_{i + 1}],
\label{eq:s-decomposition-thj}
\end{align}
where $|\mathcal{J}|$ is the number of elements in the set
\begin{align}
\label{eq:num-elements-i-thj}
\mathcal{I} := \left\{ (o, d) : \exists \alpha \in \R \text{ such that } o = \mathcal{O}(y'(\alpha)), d = \Delta(y'(\alpha)),  \thj \in \mathcal{M}(y'(\alpha)) \right\}.
\end{align}
$\mathcal{I}$ is the set of possible orders and signs of the changepoints that can be obtained via a perturbation of $y$ that yields a changepoint at $\thj$.
\end{prop}

\begin{prop}
\label{prop:efficient-intervals-thj}
$\bigcup_{i\in\mathcal{J}} [a_{i}, a_{i+1}]$ defined in \eqref{eq:s-decomposition-thj} can be efficiently computed. 
\end{prop}

We outline the proof for Proposition~\ref{prop:efficient-intervals-thj} here. We first run $k$-step binary segmentation on the data $y$ in order to obtain estimated changepoints $\mathcal{M}(y)$, orders $\mathcal{O}(y)$, and signs $\Delta(y)$. We then apply the first statement in Proposition~\ref{prop:s-decomposition-thj} to obtain an interval $[a_{0}, a_{1}]$. Since $[a_{0}, a_{1}] \subset \mathcal{S}$, we set $\mathcal{J} = \{0\}$. 
Next, for some small positive value of $\eta$,  we apply the first statement of Proposition~\ref{prop:s-decomposition-thj} with $m = \mathcal{M}(y'(a_1+\eta))$, $o = \mathcal{O}(y'(a_1+\eta))$, and $d = \Delta(y'(a_1+\eta))$ to identify the interval $[a_1, a_2]$. We then check whether $\thj \in \mathcal{M}(y'(a_1+\eta))$; if so, then $[a_1, a_2] \subset \mathcal{S}$, and we update $\mathcal{J}$ to $\mathcal{J}\cup \{1\}$. We continue in this vein, much as we did in Section~\ref{sec:bs-characterization}, to obtain the full set $\mathcal{J}$. 

In fact, when characterizing the set $\mathcal{S} = \{ \phi : \thj \in \mathcal{M}(y'(\phi)) \}$, this procedure can be sped up. For some positive integer $q$, consider the interval in $\phi$ such that $q$-step binary segmentation applied to $y'(\phi)$ yields estimated changepoints with locations $m$, orders $o$, and signs $d$,
\begin{align}
\label{eq:j-step}
\{\phi : \mathcal{M}_{q}(y'(\phi)) = m, \mathcal{O}_{q}(y'(\phi)) = o, \Delta_{q}(y'(\phi)) = d \},
\end{align}
 where the subscripts indicate that we have used $q$-step binary segmentation as opposed to $k$-step binary segmentation. Note that if $q \ll k$, then the interval in \eqref{eq:j-step} can be computed
much more quickly than the interval in the first statement of Proposition~\ref{prop:efficient-intervals-thj}, which is obtained using $k$-step binary segmentation. 

Now, recall that $\thj$ is the $j$th estimated changepoint resulting from binary segmentation on the data $y$. Suppose that $j<k$.  We first run $j$-step binary segmentation on $y$ in order to obtain estimated changepoints $\mathcal{M}_{j}(y)$, orders $\mathcal{O}_{j}(y)$, and signs $\Delta_{j}(y)$. Then we can identify an interval $[a_0, a_1] \subset \mathcal{S}$ by applying \eqref{eq:j-step} with $m = \mathcal{M}_{j}(y)$, $o = \mathcal{O}_{j}(y)$, and $d = \Delta_{j}(y)$. 
This leads to substantial computational speed-ups if $j \ll k$. 
Next, suppose that $\thj$ is the $l$th estimated changepoint resulting from $k$-step binary segmentation applied to $y'(a_1 + \eta)$, for $l  < k$. Once again, we can identify an interval $[a_1,a_2] \subset \mathcal{S}$ by applying \eqref{eq:j-step} with $m = \mathcal{M}_{l}(y'(a_1 + \eta))$, $o = \mathcal{O}_{l}(y'(a_1 + \eta))$, and $d = \Delta_{l}(y'(a_1 + \eta))$. By contrast, if $\thj \notin \mathcal{M}_{k}(y'(a_1 + \eta))$ or if $\thj$ is the $k$th estimated changepoint on the data $y'(a_1+\eta)$, then we must identify intervals using the first statement of Proposition~\ref{prop:s-decomposition-thj}.

\section{Details related to Section~\ref{sec:l0-characterization}}

\subsection{Proof of Theorem~\ref{thm:2d-pruning}}
\label{sec:proof-thm}

To compute $\cost(y_{1:s}'({\phi}); u)$ for  $s \in \{ \thi{j-1} +1, \ldots, \thi{j}\}$, 
%we wish to calculate the cost associated with the data $y_{1:s}'(\phi)$, given that the mean at the $s$th timepoint is $u$. %There are two possibilities: 
%To do this, 
we will introduce a set of functions $\C_s$; each function in the set will correspond to a possible configuration for the changepoints preceding the $s$th timepoint.  Then, $\cost(y_{1:s}'(\phi);u) = \min_{f \in \C_s} f(u, \phi)$. Importantly, we will  construct the set $\C_s$ in such a way  that its size grows linearly, rather than exponentially, in $s$. %the size of the set of values that $s$ can take. 

To begin,  we let $\C_{\thi{j-1}} = \{ \cost(y_{1:\thi{j-1}}; u) \}$ be a set containing a single function, $\cost(y_{1:\thi{j-1}}; u)$,  which can be obtained by applying \eqref{eq:functional-updates} for $s = 1, \ldots, \thi{j-1}$. 
To obtain the set $\C_{\thi{j-1}+1}$, we must update $\C_{\thi{j-1}}$ to allow for the following two possibilities:
\begin{enumerate}
 \item \emph{There is no changepoint at the $(\thi{j-1})$th timepoint.} In this case, the cost is  \[ \cost(y_{1:\thi{j-1}}; u) + \frac12\left(y_{\thi{j-1}+1}'({\phi}) - u\right)^{2}.\]
 \item \emph{There is a changepoint at the $(\thi{j-1})$th timepoint.} This incurs a penalty of $\lambda$, and leads to a cost of 
 \[ 
 \min_{u'} \left\{ \cost(y_{1:\thi{j-1}}; u') \right\}  + \frac12\left(y_{\thi{j-1}+1}'({\phi}) - u\right)^{2} + \lambda.
 \]
\end{enumerate}

Therefore, $\cost(y_{1:(\thi{j-1}+1 )}'(\phi); u) = \min_{f \in \C_{\thi{j-1}+1}} f(u, \phi)$, where 
$$\C_{\thi{j-1}+1} = \left\{ \cost(y_{1:\thi{j-1}}; u) + \frac12 (y_{\thi{j-1}+1}'({\phi}) - u)^{2}, \;  \min_{u'}  \left\{ \cost(y_{1:\thi{j-1}}; u') \right\} + \frac12 (y_{\thi{j-1}+1}'({\phi}) - u)^{2} + \lambda \right\}. $$

Continuing on to the next timepoint, we can see that $\cost(y_{1:(\thi{j-1}+2 )}'(\phi); u) = \min_{f \in \C_{\thi{j-1}+2}} f(u, \phi)$, where
\begin{align}
\C_{\thi{j-1}+2} = \Bigg\{ & \cost(y_{1:\thi{j-1}}; u) + \frac12 (y_{\thi{j-1}+1}'({\phi}) - u)^{2} + \frac12 (y_{\thi{j-1}+2}'({\phi}) - u)^{2}, \label{eq:1}\\  & \min_{u'} \left\{  \cost(y_{1:\thi{j-1}}; u') \right\} + \frac12 (y_{\thi{j-1}+1}'({\phi}) - u)^{2} + \lambda + \frac12 (y_{\thi{j-1}+2}'({\phi}) - u)^{2}, \label{eq:2}\\
& \min_{u''} \left\{ \cost(y_{1:\thi{j-1}}; u'') + \frac12 (y_{\thi{j-1}+1}'({\phi}) - u'')^{2} \right\} +  \frac12 (y_{\thi{j-1}+2}'({\phi}) - u)^{2} + \lambda, \label{eq:3}\\
&  \min_{u'}  \left\{ \cost(y_{1:\thi{j-1}}; u') \right\}  + \min_{u''} \left\{ \frac12 (y_{\thi{j-1}+1}'({\phi}) - u'')^{2} + \lambda \right\} 
  +  \frac12 (y_{\thi{j-1}+2}'({\phi}) - u)^{2} + \lambda \label{eq:4}
\Bigg\}. 
\end{align}
Here, \eqref{eq:1} corresponds to no changepoint at either $\thi{j-1}$ or $\thi{j-1}+1$, \eqref{eq:2} corresponds to a changepoint at $\thi{j-1}$, \eqref{eq:3} corresponds to a changepoint at $\thi{j-1} + 1$, and \eqref{eq:4} corresponds to  changepoints at $\thi{j-1}$ and $\thi{j-1}+1$. 
We could continue along this vein to create the sets $\C_{\thi{j-1}+3},\ldots,\C_{\thi{j}}$, but the number of functions in the sets would scale exponentially, making computations intractable. 
 Instead, we notice that we really care about the \emph{minimum} of the functions in each set, as a function of $u$ and $\phi$; furthermore, 
 since \eqref{eq:3} and \eqref{eq:4} are of the form $h(\phi) + \frac12 (y_{\thi{j-1}+2}'({\phi}) - u)^{2} + \lambda$, their minimum takes the form  
 {\small
 \begin{align}
&  \min\left\{ \min_{u''} \left\{ \cost(y_{1:\thi{j-1}}; u'') + \frac12 (y_{\thi{j-1}+1}'({\phi}) - u'')^{2} \right\} , \; \min_{u'}  \left\{ \cost(y_{1:\thi{j-1}}; u') \right\}   + \min_{u''} \left\{  \frac12 (y_{\thi{j-1}+1}'({\phi}) - u'')^{2}  + \lambda \right\} \right\}  \nonumber \\
 &  \hspace{10mm} +  \frac12 (y_{\thi{j-1}+2}'({\phi}) - u)^{2} + \lambda. \label{eq:min}
 \end{align}}
 Thus, it is not necessary for us to keep track of \eqref{eq:3} and \eqref{eq:4}; we can just keep track of \eqref{eq:min} instead. Using this insight, as $s$ increases by one, the set  $\C_s$ will increase by just one function, rather than increasing exponentially. 
 Importantly, \eqref{eq:min} is a piecewise quadratic function of $\phi$, plus a quadratic function of $\phi$ and $u$; therefore, it can be efficiently calculated and stored
  using ideas 
 from \cite{rigaill2015pruned} and \cite{maidstone2017optimal}.    

We now summarize the overall procedure. For  $s = \thi{j - 1}+1, \ldots, \thj$, 
  we update the set $\C_{s}$ as 
\begin{align}
\C_{s} = \left\{ f(u, \phi) + \frac12 (y_{s}'({\phi}) - u)^{2} : f \in \C_{s - 1} \cup \{h_{s}(\phi) \}		\right\},
\label{eq:collection}
\end{align}
where 
\begin{align}
h_{s}(\phi) &= \min_{f \in \C_{s -1}} \min_{u'} f(u', \phi) + \lambda.
\label{eq:min-collection}
\end{align}

Furthermore, from \eqref{eq:collection}--\eqref{eq:min-collection}, the size of the set $\C_s$ increases by one as $s$ increases by one. Therefore,  computing  $\cost(y_{1:\thj}'(\phi);u)$ requires $1 + 2 + \ldots + (\thj-\thi{j-1}) = \mathcal{O}\left((\thi{j}-\thi{j-1})^2\right)$ operations in the case of  \eqref{eq:set-s}.

\subsection{Characterization of \eqref{eq:set-s-thj}}
\label{sec:set-s-thj-l0}

In this section, we show that we can characterize the set $\mathcal{S} \equiv \{ \phi: \thj \in \mathcal{M}(y'(\phi)) \}$ for changepoints estimated via $\ell_{0}$ segmentation. 
For $\mathcal{S}$ defined in \eqref{eq:set-s-thj}, $\phi \in \mathcal{S}$ if and only if the cost of segmenting $y'_{1:T}(\phi)$ with a changepoint at $\thj$,
\begin{align}
\tilde{C}(\phi) &= \min_{u}\left\{ \cost(y_{1:\thj}'({\phi}); u)\right\} + \min_{u} \left\{\cost(y_{T:(\thj+1)}'({\phi}); u) \right\}+ \lambda, \label{eq:Ctilde}
\end{align}
is no greater than the cost of segmenting $y_{1:T}'(\phi)$ with no changepoint at $\thj$,
\begin{align}
\tilde{C}'(\phi) &= \min_{u} \left\{ \cost(y_{1:\thj}'({\phi}); u) + \cost(y_{T:(\thj+1)}'({\phi}); u)\right\},
\label{eq:Ctildeprime}
\end{align}
where $\cost(y_{1:s}; u)$ is defined in \eqref{eq:functional-updates}. Therefore, $\mathcal{S} = \{ \phi :  \thj \in \mathcal{M}(y'(\phi))\} = \{\phi : \tilde{C}(\phi) \leq \tilde{C}'(\phi)\}$. We note that \eqref{eq:Ctilde} and \eqref{eq:Ctildeprime} are identical to \eqref{eq:c-phi-thj} and \eqref{eq:c-phi-neg-thj} defined in Section~\ref{sec:l0-characterization}, except here the contrast $\nu$ is defined in \eqref{eq:nu-window}, whereas in Section~\ref{sec:l0-characterization} it is defined in \eqref{eq:nu}. Therefore, we can compute $\mathcal{S}$ using a slightly modified version of the procedure of Section~\ref{sec:l0-characterization}. Section~\ref{sec:2d-fun-prune-example} of the Supplementary Materials illustrates the details on a small example.  

We also note that computing  $\cost(y_{1:\thj}'(\phi);u)$ requires $1 + 2 + \ldots + h= \mathcal{O}(h^2)$ operations in the case of  \eqref{eq:set-s-thj}. Timing results are presented in Section~\ref{sec:appendix-timing} of the Supplementary Materials.

\subsection{An illustration of the procedure of Section~\ref{sec:set-s-thj-l0}}
\label{sec:2d-fun-prune-example}

To better grasp the procedure described in Section~\ref{sec:set-s-thj-l0} of the Supplementary Materials to characterize the set $\mathcal{S} = \{ \phi : \thj \in \mathcal{M}(y'(\phi))\}$ in \eqref{eq:set-s-thj} for $\ell_{0}$ segmentation, in this section we work through a simple example. Suppose we observe $y = [1, 1, 1, 2, 2, 2]$, and estimate a changepoint at $\hat\tau = 3$ by solving \eqref{eq:l0-mean-rss} with $\lambda = \frac12$.

In this example, we take $h = 2$, and use the simplified perturbation model 
\begin{align}
y_{t}'({\phi}) &= 
\begin{cases}
 y_{t} & t=1, 6, \\
 y_{t} + \phi & t = 2, 3, \\
 y_{t} - \phi & t = 4, 5.
\end{cases}
\label{eq:simplified-yphi}
\end{align}

We wish to ultimately compute $\C_{3}$, so we begin with $\C_{1} =  \{ \cost(y_{1}; u) \}$,
\begin{align*} 
\cost(y_{1}; u) = \frac12 (1 - u)^{2},
\end{align*}
and repeatedly use \eqref{eq:collection} and \eqref{eq:min-collection} to obtain $\C_{2}$ from $\C_{1}$ and $\C_{3}$ from $\C_{2}$. 

$\C_{2}$ contains two functions: the first function represents the cost of segmenting $[1, 1 + \phi]$ with zero changepoints and where the mean $\mu_{2} = u$; the second function represents the cost of segmenting $[1, 1 + \phi]$ with a changepoint at timepoint 1, and where the mean $\mu_{2} = u$. By \eqref{eq:collection}, this is simply 
\[\C_{2} = \left\{ \frac12 (1 - u)^{2} + \frac12 (1 + \phi - u)^{2}, h_{2}(u, \phi) + \frac12 (1 + \phi - u)^{2} \right\},\] where 
\begin{align*}
h_{2}(u, \phi) = \min_{u'} \cost(y_{1}; u') + \lambda = \min_{u'} \frac12 (1 - u')^{2} + \frac12 = \frac12.
\end{align*}
More explicitly, 
\begin{align*}
\C_{2}&= \left\{ \frac12 (1 - u)^{2} + \frac12 (y_{2}'({\phi}) - u)^{2}, \frac12 + \frac12 (y_{2}'({\phi}) - u)^{2} \right\} \\
&= \left\{u^{2} -2u -u\phi + \frac12 \phi^{2} + \phi + 1, \frac12 u^2 -u-u\phi+\frac12\phi^2+\phi + 1 \right\}.
\end{align*} 
To compute $\C_{3}$, we first calculate the minimum (corresponding to a changepoint at timepoint 2) 
\begin{align*}
h_{3}(u, \phi)= \min_{f \in \C_{2}} \min_{u'}  f(u', \phi) + \lambda  = 
\begin{cases}
1 & \phi < -\sqrt{2} \\
 \frac14 \phi^{2} + \frac12 & -\sqrt{2} \leq \phi \leq \sqrt{2} \\
1 & \phi > \sqrt{2}
\end{cases},
\end{align*}
and add the perturbed data point, $1 + \phi$, to obtain $\C_{3}= \{ q_{1}(u, \phi), q_{2}(u, \phi), q_{3}(u, \phi)\}$, where 
\begin{align*}
q_{1}(u, \phi) &=  1.5u^2 -3u-2u\phi+\phi^2+2\phi+1.5, \\
q_{2}(u, \phi) &= u^2 -2u-2u\phi+\phi^2+2\phi+1.5, \\
q_{3}(u, \phi) &= \begin{cases}
0.5u^2 -u-u\phi+0.5\phi^2+\phi +1.5 & \phi < -\sqrt{2} \\
0.5u^2 -u-u\phi+0.75\phi^2+\phi+1 & -\sqrt{2} \leq \phi \leq \sqrt{2} \\
0.5u^2 -u-u\phi+0.5\phi^2+\phi+1.5 & \phi > \sqrt{2}
\end{cases}
\end{align*}
For any $u$ and $\phi$, the optimal cost of segmenting $y_{1:3}'({\phi})$ is given as $\cost(y_{1:3}'({\phi}); u) = \min_{f\in\C_{3}} f(u, \phi)$. 

Applying similar steps in the reverse direction from timepoint 6 to timepoint 4, gives 

\[ \cost(y_{6:4}'({\phi}); u) = \min\{f_{1}(u, \phi), f_{2}(u, \phi), f_{3}(u, \phi)\}, \] where
\begin{align*}
f_{1}(u, \phi) &= 1.5u^2 -6u+2u\phi+\phi^2-4\phi+6, \\
f_{2}(u, \phi) &= u^2 -4u+2u\phi+\phi^2-4\phi+4.5, \text{ and } \\ 
f_{3}(u, \phi) &= 
\begin{cases}
0.5u^2 -2u+u\phi+0.5\phi^2-2\phi+3 & \phi < -\sqrt{2} \\
0.5u^2 -2u+u\phi+0.75\phi^2-2\phi+2.5 & -\sqrt{2} \leq \phi \leq \sqrt{2} \\
0.5u^2 -2u+u\phi+0.5\phi^2-2\phi+3 & \phi > \sqrt{2}
\end{cases}.
\end{align*}

\begin{figure}
\begin{center}
\includegraphics[width=\textwidth]{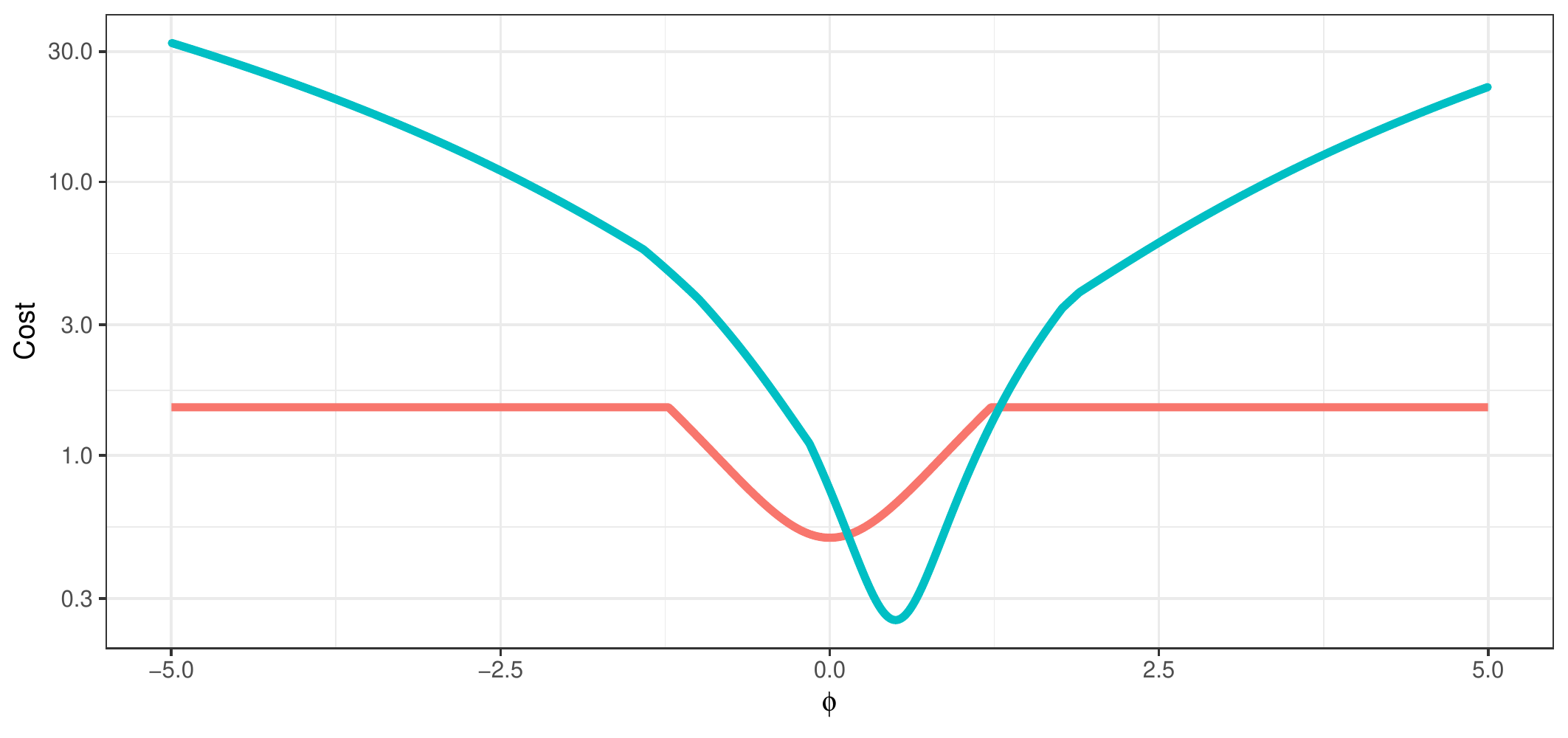}
\caption{Optimal cost of segmenting $y'({\phi})$ as a function of $\phi$, in the example in Section~\ref{sec:2d-fun-prune-example} of the Supplementary Materials. $\tilde{C}(\phi)$ is the optimal cost of segmenting $y'(\phi)$ as a function of $\phi$ given that there is a changepoint at $\hat\tau = 3$ (red). $\tilde{C}'(\phi)$ is the optimal cost of segmenting $y'(\phi)$ given that there is no changepoint at $\hat\tau = 3$ (blue).}
\label{fig:opt-cost-example}
\end{center}
\end{figure}

$\tilde{C}(\phi)$ and $\tilde{C}'(\phi)$, defined in \eqref{eq:Ctilde} and \eqref{eq:Ctildeprime}, are calculated as
\begin{align*}
\tilde{C}(\phi) &= \min_{u} \cost(y_{1:3}'({\phi}); u) + \min_{u} \cost(y_{6:4}'({\phi}); u) + \lambda 
=
\begin{cases}
\frac32 & \phi < -\sqrt{\frac32}  \\
\frac23 \phi^{2} +  \frac12 & -\sqrt{\frac32} \leq \phi \leq \sqrt{\frac32}  \\
\frac32 & \phi > \sqrt{\frac32}\\
\end{cases},
\end{align*}
and
\begin{align*}
\tilde{C}'(\phi) &= \min_{u} \left\{ \cost(y_{1:3}'({\phi}); u) + \cost(y_{6:4}'({\phi})s; u)\right\} 
=
\begin{cases}
\phi^{2} -\phi +  2.25 & \phi < -1.41421  \\
1.5\phi^{2} -\phi + 1.25 & -1.41421 \leq \phi \leq  -1  \\
1.625\phi^{2} -1.25\phi +  0.875 &  -1 \leq \phi \leq -0.1547  \\
2\phi^{2} -2\phi +  0.75 &  -0.1547 \leq \phi \leq 1.76619  \\
1.375\phi^{2} +1.375\phi + 2.25 & 1.76619 \leq \phi \leq 1.89681  \\
\phi^{2} -\phi +  2.25 & \phi > 1.89681 
\end{cases}.
\end{align*}
To determine $\mathcal S$, we recall from Section~\ref{sec:set-s-thj-l0} that $\mathcal{S} = \{ \phi : \tilde{C}(\phi) \leq \tilde{C}'(\phi)\}$. Therefore, we take the minimum
\[
\min\left\{ \tilde{C}(\phi), \tilde{C}'(\phi)\right\} = 
\left\{
\begin{aligned}
&1.5 &  \phi < -1.22474 \quad & \text{Minimizer: } \tilde{C}(\phi)  \\
&\frac23 \phi +\frac12 & -1.22474 \leq \phi \leq 0.13763 \quad& \text{Minimizer: } \tilde{C}(\phi)  \\
&2\phi^{2} -2\phi + 0.75 & 0.13763 \leq \phi \leq 1.29057 \quad& \text{Minimizer: } \tilde{C}'(\phi)  \\
& 1.5 & \phi > 1.29057 \quad& \text{Minimizer: } \tilde{C}(\phi)  \\
\end{aligned}
\right.
\]
and for each point $\phi$ track whether $\tilde{C}(\phi)$ or $\tilde{C}'(\phi)$ minimized the objective. Therefore,  $\mathcal S = (-\infty, 0.13763] \cup [1.29057, \infty)$. Figure~\ref{fig:opt-cost-example} shows $\tilde{C}(\phi)$ and $\tilde{C}'(\phi)$.

\subsection{Timing results for computing the set $\mathcal{S}$ defined in \eqref{eq:set-s-thj}}

\label{sec:appendix-timing}

In this section, we investigate the claim of Section~\ref{sec:set-s-thj-l0} of the Supplementary Materials, that computing the set $\mathcal{S}$ defined in \eqref{eq:set-s-thj} in the case of $\ell_{0}$ segmentation requires 
$\mathcal{O}(h^2)$ computations, where $h$ is the window size that appears in \eqref{eq:nu-window}. 

\begin{figure}
\begin{center}
\includegraphics[width=\textwidth]{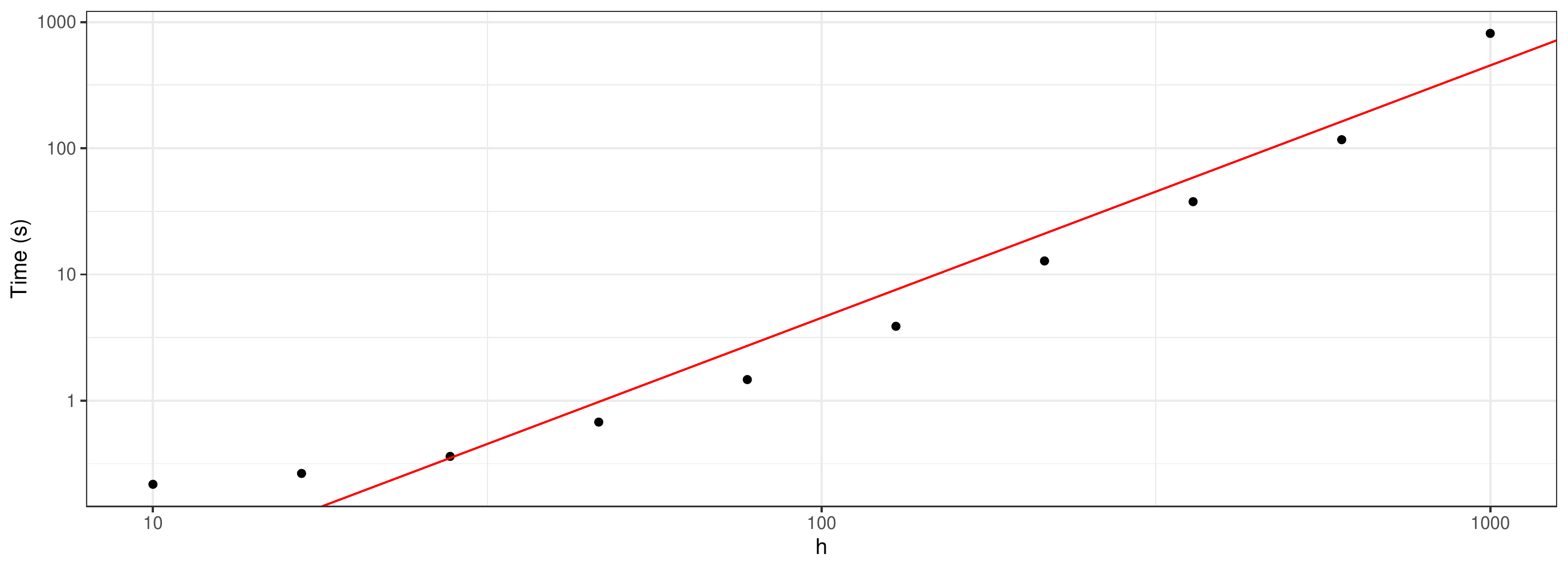}
\caption{Average time, in seconds, to compute the set $\mathcal{S}$ in \eqref{eq:set-s-thj},  as a function of the window size $h$ on 50 replicated datasets. Both axes are displayed on the log scale. The function $\mathrm{time} = e^{-3.3} h^2$ (red) is displayed for reference. Details are provided in Section~\ref{sec:appendix-timing} of the Supplementary Materials.}
\label{fig:running-time-example}
\end{center}
\end{figure}

Figure~\ref{fig:running-time-example} displays the average running time over 50 replicate datasets as a function of the window size, $h$,  on a simulated dataset of $2000$ timepoints, which contains a single changepoint at the $1000$th timepoint.  We see that the   running time is, in fact, approximately quadratic in the window size.

\section{Efficient analytical characterization of \eqref{eq:set-s} and \eqref{eq:set-s-thj} for the fused lasso and the generalized lasso} 

\subsection{Fused Lasso}
\label{sec:fused-lasso-characterization}

The fused lasso problem \eqref{eq:l1-mean-rss} can be reformulated as the regression problem
\begin{align}
\minimize{\beta\in\R^{T}}{|| y - \bm{X}\beta||_{2}^{2} + \lambda || \beta ||_{1}},
\label{eq:lasso-1d}
\end{align}
for a $T \times T$ matrix $\bm{X}$ whose $j$th row contains $j$ ones followed by $T - j$ zeros. \eqref{eq:l1-mean-rss} and \eqref{eq:lasso-1d} are equivalent in the sense that $\hat\beta_{t} = \hat\mu_{t} - \hat\mu_{t-1}$ for $t = 2, \ldots, T$ and $\hat\beta_{1} = \hat\mu_{1}$. 

\cite{lee2016exact} show that the set of $y$ for which the lasso \eqref{eq:lasso-1d} results in a given set of selected variables and signs can be written as the polyhedral set $\{y: \bm{A}y \leq b\}$ for a $T \times T$ matrix $\bm{A}$ and a $T$-vector $b$. $\bm{A}$ and $b$ have explicit formulas depending only on the selected variables and coefficient signs. Therefore, \cite{lee2016exact}  are able to compute $p$-values for the null hypothesis that the estimated coefficients are zero conditional on the selected variables, the signs of the estimated coefficients, and nuisance parameters. 

To avoid conditioning on the signs of the estimated coefficients, we slightly modify the arguments outlined in Section~\ref{sec:bs-characterization}. In the following propositions, Propositions~\ref{prop:s-decomposition-lasso} and \ref{prop:efficient-intervals-lasso}, we modify Propositions~\ref{prop:s-decomposition}
and \ref{prop:efficient-intervals} for $\mathcal{S} = \{ \phi : \mathrm{supp}(\hat\beta(y'(\phi))) = \mathrm{supp}(\hat\beta(y)) \}$, where $\mathrm{supp}(\hat\beta(y))$ denotes the set of selected variables obtained from solving \eqref{eq:lasso-1d} with data $y$.

\begin{prop}
\label{prop:s-decomposition-lasso}
The set $\{ \phi : \mathrm{supp}(\hat\beta(y'(\phi))) = m, \mathrm{sign}(\hat\beta(y'(\phi))) = d \}$ is an interval.
Furthermore, the set $\mathcal{S} = \{ \phi : \mathrm{supp}(\hat\beta(y'(\phi))) = \mathrm{supp}(\hat\beta(y)) \}$ can be written as the union of intervals,
\begin{align}
\mathcal{S} = \{ \phi: \mathrm{supp}(\hat\beta(y'(\phi))) = \mathrm{supp}(\hat\beta(y)) \} &= \bigcup_{i \in \mathcal{J}} [a_{i}, a_{i + 1}],
\label{eq:s-decomposition-lasso}
\end{align}
where $|\mathcal{J}|$ is the number of elements in the set
\begin{align}
\label{eq:num-elements-i-lasso}
\mathcal{I} := \left\{ d : \exists \alpha \in \R \text{ such that } d = \mathrm{sign}(\hat\beta(y'(\alpha))), \mathrm{supp}(\hat\beta(y)) = \mathrm{supp}(\hat\beta(y'(\alpha)))  \right\}.
\end{align}
$\mathcal{I}$ is the set of possible coefficient signs that can be obtained via a perturbation of $y$ that yields the same non-zero coefficients as $\hat\beta(y)$.
\end{prop}

\begin{prop}
\label{prop:efficient-intervals-lasso}
$\bigcup_{i\in\mathcal{J}} [a_{i}, a_{i+1}]$ defined in \eqref{eq:s-decomposition-lasso} can be efficiently computed. 
\end{prop}

Now, we outline the proof for Proposition~\ref{prop:efficient-intervals-lasso}. We first solve \eqref{eq:lasso-1d} on the data $y$ in order to obtain $\mathrm{supp}(\hat\beta(y))$ and $\mathrm{sign}(\hat\beta(y))$. We then apply the first statement in Proposition~\ref{prop:s-decomposition-lasso} to obtain an interval $[a_{0}, a_{1}] \subset \mathcal{S}$. We initialize $\mathcal{J} = \{0\}$.  
Next, for some small positive value of $\eta$,  we apply the first statement of Proposition~\ref{prop:s-decomposition-lasso} with $m = \mathrm{supp}(\hat\beta(y'(a_{1} + \eta)))$ and $d =\mathrm{sign}(\hat\beta(y'(a_{1} + \eta))) $ to identify the interval $[a_1, a_2]$. If $\mathrm{supp}(\hat\beta(y)) = \mathrm{supp}(\hat\beta(y'(a_1+\eta)))$ we set $\mathcal{J}$ to $\mathcal{J} \cup \{1\}$. We continue in this vein, much as we did in Section~\ref{sec:bs-characterization}, to obtain the full set $\mathcal{J}$.

\subsection{Generalized lasso} 
\label{sec:generalized-lasso}

In this section, we show that we can use the tools from Section~\ref{sec:bs-characterization} to characterize the selection event of the generalized lasso \citep{tibshirani2011solution}, which is the solution to the optimization problem 
\begin{align}
\minimize{\beta\in\R^{T}}{|| y - \beta||_{2}^{2} + \lambda || \bm{D} \beta ||_{1}}.
\label{eq:generalized-lasso}
\end{align}
For general $\bm{D}$, \eqref{eq:generalized-lasso} cannot be rewritten in the form of \eqref{eq:lasso-1d}, and so existing machinery for selective inference for the lasso cannot be applied. Nonetheless, by also conditioning on the order that variables enter the model, \cite{hyun2016exact} show that the selection event of the generalized lasso is polyhedral. Therefore, an extension of the ideas in Section~\ref{sec:fused-lasso-characterization} could be applied in order to conduct selective inference using a larger conditioning set.

\section{Timing results for estimating changepoints and computing $p$-values}
\label{sec:timing-relative}

\begin{figure}
\begin{center}
\includegraphics[width=\textwidth]{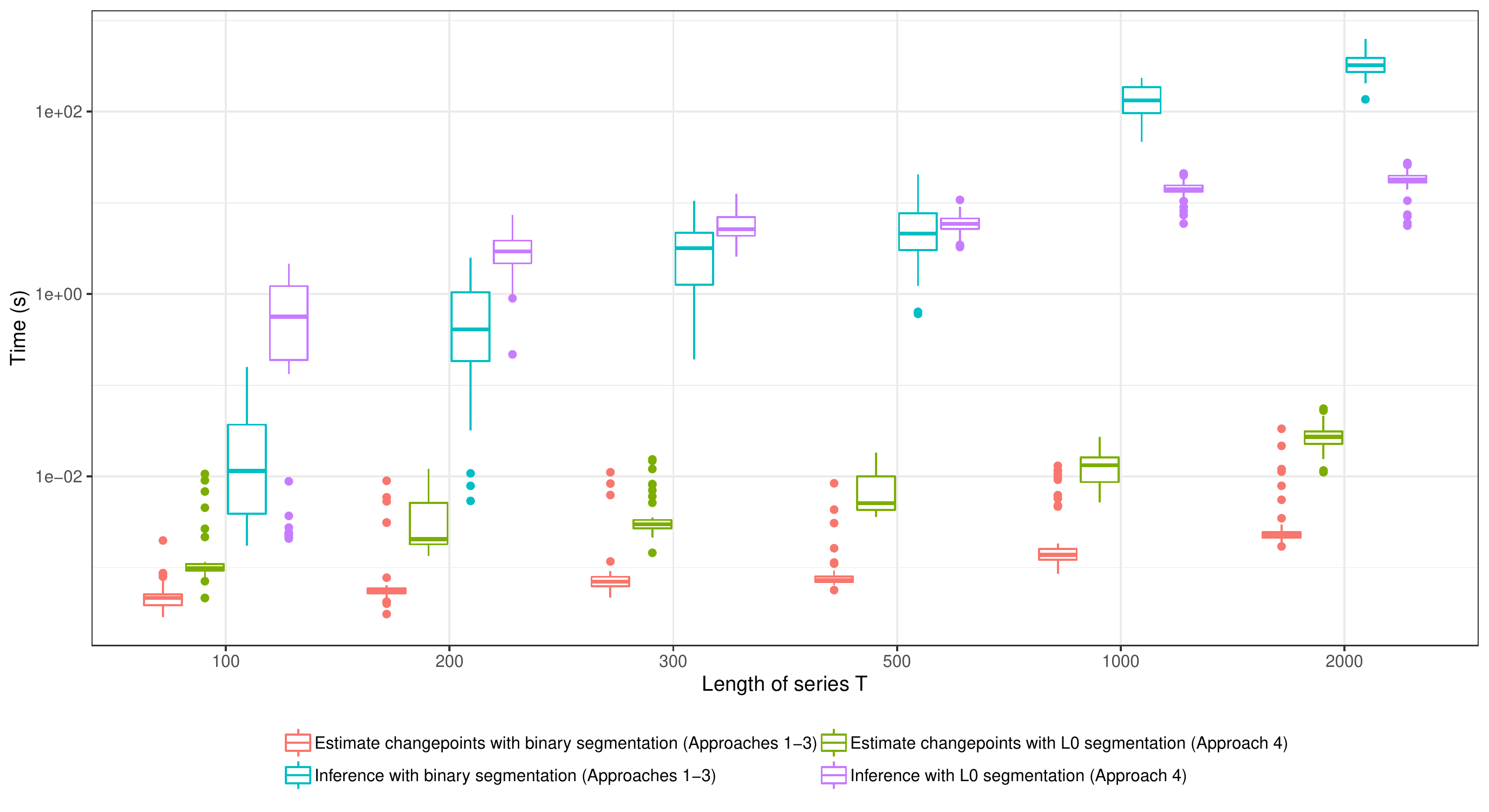}
\caption{Computational cost of Approaches 1--4 defined in Section~\ref{sec:exp-intro}. 50 replicate datasets are simulated according to model \eqref{eq:obs-model} with $\sigma^{2} = 1$ and with $K = 10 \floor{ \log_{10}(T) } \text{ } $ changepoints sampled without replacement from $\{1, \ldots, T\}$. At each changepoint the absolute difference in mean, $| \mu_{\tau_{j} + 1} - \mu_{\tau_{j}}|$, is 1.5. Details are provided in Section~\ref{sec:timing-relative} of the Supplementary Materials.}
\label{fig:relative-running-time-example}
\end{center}
\end{figure}

In this section, we present timing results for estimating changepoints and computing $p$-values.
Figure~\ref{fig:relative-running-time-example} displays the running time, computed on a MacBook Pro with a 2.5 GHz Intel Core i7 processor, for estimating changepoints and calculating $p$-values for Approaches 1--4 defined in Section~\ref{sec:exp-intro}. We take $\lambda = \log(T)$ for $\ell_{0}$ segmentation and use $\max(\hat{K}, 1)-$step binary segmentation for $\hat{K}$ equal to the number of estimated changepoints from $\ell_{0}$ segmentation. Fifty replicate datasets are simulated according to model \eqref{eq:obs-model} with $\sigma^{2} = 1$, and with $K = 10 \floor{ \log_{10}(T) } \text{ } $ changepoints sampled without replacement from the set $\{1, \ldots, T\}$. At each changepoint, the absolute difference in mean is $| \mu_{\tau_{j} + 1} - \mu_{\tau_{j}}| = 1.5$. Our implementations of Approaches~1--3 approximate the set $\mathcal{S}$ with $\tilde{\mathcal{S}}$ as described in Proposition~\ref{prop:early-stopping}; we take $|a_{-r}| = |a_{r'+1}| = \max( 10 \sigma||\nu||_{2}, |\nu^{\top}y| )$.

Estimating changepoints with binary and $\ell_{0}$ segmentation is very fast (under 0.06 seconds for all values of $T$ considered). On the other hand, inference is much more costly for all approaches. In particular, we note that Approach~4 is almost an order of magnitude faster than Approaches~1--3 for larger values of $T$. Approach~3 can be sped up using the idea presented in Section~\ref{sec:basic-set-s-character-bs-2}. 

\section{Type I error control and power for unknown $\sigma$}
\label{sec:performance-unknown-sigma}

Recall that $\sigma^2$ denotes the true error variance, i.e. $Y_t \sim N(\mu_t, \sigma^2)$. 
We can think of the $p$-value in \eqref{eq:p2}  as a function of $Y$ and $\sigma$, as follows: 
$$p(Y, \sigma) = Pr(|\nu^\top Y| \geq |\nu^\top y| \;\; |  \;\; \mathcal{M}(Y)=\mathcal{M}(y), \Pi^\perp_\nu Y = \Pi^\perp_\nu y).$$
%In the manuscript, we assume that $\hat\sigma=\sigma$.
Then, conditional on the selection event, we have that
$$p(Y, \sigma) \sim \mathrm{Unif}(0, 1)$$
under $H_0: \nu^\top \mu =0$. 

Now suppose that we have a consistent estimator of $\sigma$, i.e. $\hat\sigma \rightarrow^p \sigma$. Then,
%$$p(Y, \hat\sigma) = p(Y, \sigma) \cdot \frac{p(Y, \hat\sigma)}{p(Y, \sigma)}}.$$
%Furthermore, 
 since $p(Y, \sigma)$ is continuous in $\sigma$, we have that $\frac{p(Y, \hat\sigma)}{p(Y,\sigma)} \rightarrow^p 1$. % by the continuous mapping theorem. 
This means that under $H_0: \nu^\top \mu=0$,  
$$p(Y, \hat\sigma) = p(Y, \sigma) \cdot \frac{p(Y, \hat\sigma)}{p(Y, \sigma)} \rightarrow^d \mathrm{Unif}(0,1).$$
%due to Slutsky's Theorem. 
 So  if we plug in  a consistent estimator $\hat\sigma$ of $\sigma$ into the $p$-values in this paper, then they will asymptotically follow a $\mathrm{Unif}(0,1)$ distribution.

We now show that when $\sigma$ is unknown, then using a consistent estimator of $\sigma$ leads to reasonable Type I error control and substantial power under the same simulation setup described in Section~\ref{sec:experiments} of the main paper. Specifically, we estimate $\sigma$ using
 $$\hat\sigma = \frac{ \mathrm{median}_{t=2,\ldots,T} \left( | z_t - \mathrm{median}_{t=2,\ldots,T} (z_t) | \right) }{  \left( \Phi^{-1} (3/4) \right) \sqrt{2} },$$
where $z_t=y_t-y_{t-1}$, and where $\Phi$ is the cumulative density function of the $N(0,1)$ distribution.  When the number of changepoints $K$ is fixed, then as the number of timepoints $T \rightarrow \infty$, we have that $\hat\sigma \rightarrow^p \sigma$. 
Therefore, plugging $\hat\sigma$ into the selective $p$-values in the main paper will result in asymptotic Type 1 error control under  $H_0: \nu^\top \mu=0$. %, and thus 

\begin{figure}[h!]
\centering
\includegraphics[width = 0.8\textwidth]{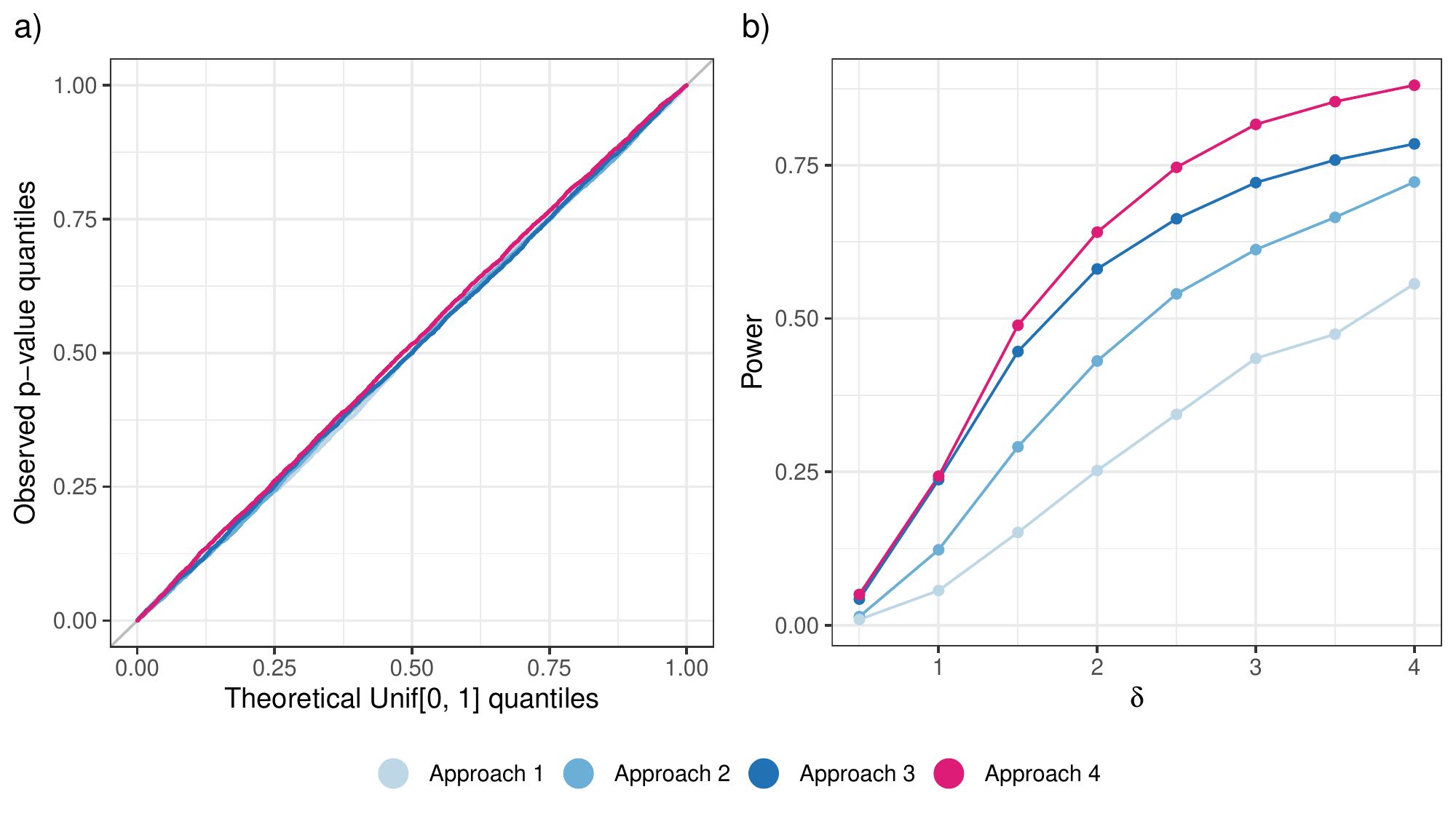}
\caption{a) Quantile-quantile plot of sample $p$-value quantiles under \eqref{eq:obs-model} with $\mu_1=\ldots=\mu_{2000}$ and a consistent estimator of $\sigma$, versus the theoretical quantiles of the $\mathrm{Unif}(0, 1)$ distribution, for Approaches 1--4 in Section~\ref{sec:exp-intro} of the main paper. b) 
 Empirical power, for Approaches 1--4 defined in Section~\ref{sec:exp-intro} of the main paper, averaged over 100 replicates, using a consistent estimator of $\sigma$.  
  }
\label{fig:est-sigma-type-i-error-power}
\end{figure}

We use $\hat\sigma$ as a plug-in value to calculate selective $p$-values for changepoints estimated via Approaches 1--4 in Section~\ref{sec:exp-intro} of the main paper. As shown in panel a) of Figure~\ref{fig:est-sigma-type-i-error-power}, this estimator yields Type I error control. Panel b) shows that this estimator results in only a very slight decrease in power relative to using the true value of $\sigma$ (see Figure~\ref{fig:l0_bs_delta_diff_seg_lengths}(a) of the main paper).

\clearpage
\bibliographystyle{apalike}
\bibliography{bib}